\documentclass[twocolumn]{aastex63}
\usepackage{graphicx}
\usepackage{float}
\usepackage[T1]{fontenc}
\usepackage{aecompl}
\usepackage[]{amsmath}
\usepackage{color}
\usepackage{lineno}
\usepackage{xspace}

\usepackage{soul}
\usepackage{booktabs} 
\usepackage{array} 
\usepackage{multirow} 
\usepackage{orcidlink}
\usepackage{tabularx}
\usepackage{booktabs}
\usepackage{rotating} 
\usepackage{amsmath}

\graphicspath{{figures/}}

\def\lsim{ \lower .75ex \hbox{$\sim$} \llap{\raise .27ex \hbox{$<$}} }

\submitjournal{ApJ}

\shorttitle{Velocity field of MW stellar halo}
\shortauthors{Li et al.}
\graphicspath{{./}{figures/}}

\begin{document}

\title{The velocity field of our Milky Way outer stellar halo based on DESI DR2}

\author[0000-0002-6469-8263]{Songting Li}
\thanks{songtingli@sjtu.edu.cn}
\affiliation{Department of Astronomy, School of Physics and Astronomy, and Key Laboratory for Particle Astrophysics and Cosmology (MOE)/Shanghai Key Laboratory for Particle Physics and Cosmology, Shanghai Jiao Tong University, Shanghai 200240, People's Republic of China}
\affiliation{Tsung-Dao Lee Institute, Shanghai Jiao Tong University, Shanghai, 201210, China}
\affiliation{State Key Laboratory of Dark Matter Physics, School of Physics and Astronomy, Shanghai Jiao Tong University, Shanghai 200240, China}
\author[0000-0002-5762-7571]{Wenting Wang}
\thanks{wenting.wang@sjtu.edu.cn}
\affiliation{Department of Astronomy, School of Physics and Astronomy, and Key Laboratory for Particle Astrophysics and Cosmology (MOE)/Shanghai Key Laboratory for Particle Physics and Cosmology, Shanghai Jiao Tong University, Shanghai 200240, People's Republic of China}
\affiliation{State Key Laboratory of Dark Matter Physics, School of Physics and Astronomy, Shanghai Jiao Tong University, Shanghai 200240, China}
\author[0000-0003-2644-135X]{Sergey E. Koposov}
\affiliation{Institute for Astronomy, University of Edinburgh, Royal Observatory, Blackford Hill, Edinburgh EH9 3HJ, UK}
\affiliation{Institute of Astronomy, University of Cambridge, Madingley Road, Cambridge CB3 0HA, UK}
\author[0000-0002-7662-5475]{Jo\~{a}o A. S. Amarante}
\affiliation{Department of Astronomy, School of Physics and Astronomy, and Key Laboratory for Particle Astrophysics and Cosmology (MOE)/Shanghai Key Laboratory for Particle Physics and Cosmology, Shanghai Jiao Tong University, Shanghai 200240, People's Republic of China}
\affiliation{State Key Laboratory of Dark Matter Physics, School of Physics and Astronomy, Shanghai Jiao Tong University, Shanghai 200240, China}
\author{Alis J. Deason}
\affiliation{Institute for Computational Cosmology, Department of Physics, Durham University, South Road, Durham DH1 3LE, UK}
\author[0000-0002-6257-2341]{Monica Valluri}
\affiliation{Department of Astronomy and Astrophysics, University of Michigan, Ann Arbor, MI, USA}
\author[0000-0002-9110-6163]{Ting S. Li}
\affiliation{David A. Dunlap Department of Astronomy \& Astrophysics, University of Toronto, 50 St George Street, Toronto ON M5S 3H4, Canada}
\author[0000-0002-5689-8791]{Amanda Bystr{\"o}m}
\affiliation{Institute for Astronomy, University of Edinburgh, Royal Observatory, Blackford Hill, Edinburgh EH9 3HJ, UK}
\author[0000-0002-2527-8899]{Mika Lambert}
\affiliation{Department of Astronomy \& Astrophysics, University of California, Santa Cruz, 1156 High Street, Santa Cruz, CA 95064, USA}
\author[0000-0002-4900-2088]{Tian Qiu}
\affiliation{Department of Astronomy, School of Physics and Astronomy, and Key Laboratory for Particle Astrophysics and Cosmology (MOE)/Shanghai Key Laboratory for Particle Physics and Cosmology, Shanghai Jiao Tong University, Shanghai 200240, People's Republic of China}
\affiliation{State Key Laboratory of Dark Matter Physics, School of Physics and Astronomy, Shanghai Jiao Tong University, Shanghai 200240, China}
\author[0000-0002-5758-150X]{Joan Najita}
\affiliation{NSF NOIRLab, 950 N. Cherry Ave., Tucson, AZ 85719, USA}
\author[0000-0003-0105-9576]{Gustavo E. Medina}
\affiliation{David A. Dunlap Department of Astronomy \& Astrophysics, University of Toronto, 50 St George Street, Toronto ON M5S 3H4, Canada}
\author[0000-0001-9852-9954]{Oleg~Y.~Gnedin}
\affiliation{Department of Astronomy and Astrophysics, University of Michigan, Ann Arbor, MI, USA}
\author[0000-0002-0740-1507]{Leandro {Beraldo e Silva}}
\affiliation{Observatório Nacional, Rio de Janeiro - RJ, 20921-400, Brasil}
\author[0000-0001-5550-2057]{Richard A. N. Brooks}
\affiliation{Department of Physics and Astronomy, University College London, London, WC1E 6BT, UK}
\author[0000-0002-7667-0081]{Raymond G. Carlberg}
\affiliation{David A. Dunlap Department of Astronomy \& Astrophysics, University of Toronto, 50 St George Street, Toronto ON M5S 3H4, Canada}
\author[0000-0003-0853-8887]{Namitha Kizhuprakkat}
\affiliation{Institute of Astronomy and Department of Physics, National Tsing Hua University, Hsinchu 30013, Taiwan}
\affiliation{Center for Informatics and Computation in Astronomy, National Tsing Hua University, Hsinchu 30013, Taiwan}
\author{Jiaxin Han}
\affiliation{Department of Astronomy, School of Physics and Astronomy, and Key Laboratory for Particle Astrophysics and Cosmology (MOE)/Shanghai Key Laboratory for Particle Physics and Cosmology, Shanghai Jiao Tong University, Shanghai 200240, People's Republic of China}
\affiliation{State Key Laboratory of Dark Matter Physics, School of Physics and Astronomy, Shanghai Jiao Tong University, Shanghai 200240, China}
\author{Jessica Nicole Aguilar}
\affiliation{Lawrence Berkeley National Laboratory, 1 Cyclotron Road, Berkeley, CA 94720, USA}
\author[0000-0001-6098-7247]{Steven Ahlen}
\affiliation{Department of Physics, Boston University, 590 Commonwealth Avenue, Boston, MA 02215 USA}
\author[0000-0001-9712-0006]{Davide Bianchi}
\affiliation{Dipartimento di Fisica ``Aldo Pontremoli'', Universit\`a degli Studi di Milano, Via Celoria 16, I-20133 Milano, Italy}
\affiliation{INAF-Osservatorio Astronomico di Brera, Via Brera 28, 20122 Milano, Italy}
\author{David Brooks}
\affiliation{Department of Physics \& Astronomy, University College London, Gower Street, London, WC1E 6BT, UK}
\author{Todd Claybaugh}
\affiliation{Lawrence Berkeley National Laboratory, 1 Cyclotron Road, Berkeley, CA 94720, USA}
\author[0000-0002-2169-0595]{Andrei Cuceu}
\affiliation{Lawrence Berkeley National Laboratory, 1 Cyclotron Road, Berkeley, CA 94720, USA}
\author[0000-0002-1769-1640]{Axel de la Macorra}
\affiliation{Instituto de F\'{\i}sica, Universidad Nacional Aut\'{o}noma de M\'{e}xico, Circuito de la Investigaci\'{o}n Cient\'{\i}fica, Ciudad Universitaria, Cd. de M\'{e}xico C.~P.~04510, M\'{e}xico}
\author{Peter Doel}
\affiliation{Department of Physics \& Astronomy, University College London, Gower Street, London, WC1E 6BT, UK}
\author[0000-0002-3033-7312]{Andreu Font-Ribera}
\affiliation{Institut de F\'{i}sica d’Altes Energies (IFAE), The Barcelona Institute of Science and Technology, Edifici Cn, Campus UAB, 08193, Bellaterra (Barcelona), Spain}
\author[0000-0002-2890-3725]{Jaime E. Forero-Romero}
\affiliation{Departamento de F\'isica, Universidad de los Andes, Cra. 1 No. 18A-10, Edificio Ip, CP 111711, Bogot\'a, Colombia}
\affiliation{Observatorio Astron\'omico, Universidad de los Andes, Cra. 1 No. 18A-10, Edificio H, CP 111711 Bogot\'a, Colombia}
\author[0000-0003-4992-7854]{Simone Ferraro}
\affiliation{Lawrence Berkeley National Laboratory, 1 Cyclotron Road, Berkeley, CA 94720, USA}
\affiliation{University of California, Berkeley, 110 Sproul Hall \#5800 Berkeley, CA 94720, USA}
\author[0000-0001-9632-0815]{Enrique Gaztañaga}
\affiliation{Institut d'Estudis Espacials de Catalunya (IEEC), c/ Esteve Terradas 1, Edifici RDIT, Campus PMT-UPC, 08860 Castelldefels, Spain}
\affiliation{Institute of Cosmology and Gravitation, University of Portsmouth, Dennis Sciama Building, Portsmouth, PO1 3FX, UK}
\author[0000-0003-3142-233X]{Satya Gontcho A Gontcho}
\affiliation{Lawrence Berkeley National Laboratory, 1 Cyclotron Road, Berkeley, CA 94720, USA}
\affiliation{University of Virginia, Department of Astronomy, Charlottesville, VA 22904, USA}
\author{Gaston Gutierrez}
\affiliation{Fermi National Accelerator Laboratory, PO Box 500, Batavia, IL 60510, USA}
\author[0000-0001-9822-6793]{Julien Guy}
\affiliation{Lawrence Berkeley National Laboratory, 1 Cyclotron Road, Berkeley, CA 94720, USA}
\author[0000-0002-6550-2023]{Klaus Honscheid}
\affiliation{Center for Cosmology and AstroParticle Physics, The Ohio State University, 191 West Woodruff Avenue, Columbus, OH 43210, USA}
\affiliation{Department of Physics, The Ohio State University, 191 West Woodruff Avenue, Columbus, OH 43210, USA}
\affiliation{The Ohio State University, Columbus, 43210 OH, USA}
\author[0000-0003-0201-5241]{Dick Joyce}
\affiliation{NSF NOIRLab, 950 N. Cherry Ave., Tucson, AZ 85719, USA}
\author[0000-0001-6356-7424]{Anthony Kremin}
\affiliation{Lawrence Berkeley National Laboratory, 1 Cyclotron Road, Berkeley, CA 94720, USA}
\author[0000-0003-1838-8528]{Martin Landriau}
\affiliation{Lawrence Berkeley National Laboratory, 1 Cyclotron Road, Berkeley, CA 94720, USA}
\author[0000-0001-7178-8868]{Laurent Le Guillou}
\affiliation{Sorbonne Universit\'{e}, CNRS/IN2P3, Laboratoire de Physique Nucl\'{e}aire et de Hautes Energies (LPNHE), FR-75005 Paris, France}
\author[0000-0002-1125-7384]{Aaron Meisner}
\affiliation{NSF NOIRLab, 950 N. Cherry Ave., Tucson, AZ 85719, USA}
\author{Ramon Miquel}
\affiliation{Institut de F\'{i}sica d’Altes Energies (IFAE), The Barcelona Institute of Science and Technology, Edifici Cn, Campus UAB, 08193, Bellaterra (Barcelona), Spain}
\affiliation{Instituci\'{o} Catalana de Recerca i Estudis Avan\c{c}ats, Passeig de Llu\'{\i}s Companys, 23, 08010 Barcelona, Spain}
\author[0000-0002-2733-4559]{John Moustakas}
\affiliation{Department of Physics and Astronomy, Siena University, 515 Loudon Road, Loudonville, NY 12211, USA}
\author[0000-0001-9070-3102]{Seshadri Nadathur}
\affiliation{Institute of Cosmology and Gravitation, University of Portsmouth, Dennis Sciama Building, Portsmouth, PO1 3FX, UK}
\author[0000-0002-0644-5727]{Will Percival}
\affiliation{Department of Physics and Astronomy, University of Waterloo, 200 University Ave W, Waterloo, ON N2L 3G1, Canada}
\affiliation{Perimeter Institute for Theoretical Physics, 31 Caroline St. North, Waterloo, ON N2L 2Y5, Canada}
\affiliation{Waterloo Centre for Astrophysics, University of Waterloo, 200 University Ave W, Waterloo, ON N2L 3G1, Canada}
\author[0000-0001-7145-8674]{Francisco Prada}
\affiliation{Instituto de Astrof\'{i}sica de Andaluc\'{i}a (CSIC), Glorieta de la Astronom\'{i}a, s/n, E-18008 Granada, Spain}
\author[0000-0001-6979-0125]{Ignasi Pérez-Ràfols}
\affiliation{Departament de F\'isica, EEBE, Universitat Polit\`ecnica de Catalunya, c/Eduard Maristany 10, 08930 Barcelona, Spain}
\author{Graziano Rossi}
\affiliation{Department of Physics and Astronomy, Sejong University, 209 Neungdong-ro, Gwangjin-gu, Seoul 05006, Republic of Korea}
\author[0000-0002-9646-8198]{Eusebio Sanchez}
\affiliation{CIEMAT, Avenida Complutense 40, E-28040 Madrid, Spain}
\author{David Schlegel}
\affiliation{Lawrence Berkeley National Laboratory, 1 Cyclotron Road, Berkeley, CA 94720, USA}
\author{Michael Schubnell}
\affiliation{Department of Physics, University of Michigan, 450 Church Street, Ann Arbor, MI 48109, USA}
\affiliation{University of Michigan, 500 S. State Street, Ann Arbor, MI 48109, USA}
\author[0000-0003-3449-8583]{Ray Sharples}
\affiliation{Centre for Advanced Instrumentation, Department of Physics, Durham University, South Road, Durham DH1 3LE, UK}
\affiliation{Institute for Computational Cosmology, Department of Physics, Durham University, South Road, Durham DH1 3LE, UK}
\author[0000-0002-3461-0320]{Joseph Harry Silber}
\affiliation{Lawrence Berkeley National Laboratory, 1 Cyclotron Road, Berkeley, CA 94720, USA}
\author{David Sprayberry}
\affiliation{NSF NOIRLab, 950 N. Cherry Ave., Tucson, AZ 85719, USA}
\author[0000-0003-1704-0781]{Gregory Tarlé}
\affiliation{University of Michigan, 500 S. State Street, Ann Arbor, MI 48109, USA}
\author{Benjamin Alan Weaver}
\affiliation{NSF NOIRLab, 950 N. Cherry Ave., Tucson, AZ 85719, USA}
\author[0000-0001-5381-4372]{Rongpu Zhou}
\affiliation{Lawrence Berkeley National Laboratory, 1 Cyclotron Road, Berkeley, CA 94720, USA}
\author[0000-0002-6684-3997]{Hu Zou}
\affiliation{National Astronomical Observatories, Chinese Academy of Sciences, A20 Datun Road, Chaoyang District, Beijing, 100101, P.~R.~China}

\clearpage

\begin{abstract}
Using 64,000 halo K giants from Dark Energy Spectroscopic Instrument (DESI) second Data Release (DR2), we decompose the Milky Way (MW) stellar halo between 3 and 160~kpc into metal-rich (MR) and metal-poor (MP) components via a Gaussian mixture model (GMM). The two populations are nearly equal in number but chemically and kinematically distinct: MR stars occupy highly radial orbits with velocity anisotropy of $\beta\approx0.94$ and metallicity dispersion $\sigma_{\rm[Fe/H]}\approx0.17$ dex, without obvious dependence on distance, and are mainly contributed by Gaia-Sausage/Enceladus (GSE) debris. MR component dominates the inner 30~kpc and re-emerges beyond 50~kpc, implying GSE debris can extend to $\sim$~70-80~kpc. MP stars exhibit a weaker radial bias of $\beta\approx0.46$, decreasing to $-$0.5 beyond 80~kpc, and with a larger metallicity dispersion of $\sigma_{\rm[Fe/H]}\approx0.46$ dex, showing signatures of multiple minor mergers. Both components exhibit net prograde rotation at $\sim$~10-30~kpc with a stronger azimuthal signal in the MP population. The non-equilibrium motions of the outer halo ($>$50~kpc) are quantified with a dipole-plus-contraction velocity field. We find that the outer halo is simultaneously contracting ($\mu_{\rm compr}=-19~\mathrm{km~s^{-1}}$, distance-independent) and subject to reflex motions ($\mu_{\rm dipole}$ increases from $-19$ to $-44~\mathrm{km~s^{-1}}$ with radius), reflecting the perturbation from the Large Magellanic Cloud (LMC). We also confirm a linear dependence of mean polar velocity for the outer stellar halo on $\mu_{\rm dipole}$, a direct consequence of the LMC and MW interaction. Our results provide a quantitative distance-resolved description of the MW's last major accretion event and its ongoing response to the first infall of the LMC.
\end{abstract}


\section{Introduction}
\label{sec:intro}

The Milky Way (MW) stellar halo is the product of galactic accretions over a long history \citep{Belokurov2018,Helmi2018,Myeong:2019aa,
Evans:2020aa,Helmi2020,Naidu2020, Naidu:2021aa,Horta2023,Horta2024_GA}. 
Rather than forming in isolation, our Galaxy has grown by repeatedly merging with smaller satellite galaxies \citep{Searle:1978aa,Johnston:1996aa, 1998ApJ...500L.149Z,Chiba_beers2000,Helmi_zeeuw2000, Bullock:2001aa,2004MNRAS.353..874M,2010MNRAS.404.1203F,2021MNRAS.505.5957B,2024OJAp....7E..23C}.

These cumulative merger events have sculpted the MW halo's global density profile and shape \citep{1987MNRAS.227P..21S,1991ApJ...375..121P,2000AJ....119.2254M,2008ApJ...680..295B,2011ApJ...731....4S,Han_stellar_halo_density_profile,joao,2024MNRAS.531.4762M,2025ApJ...985L..22N,li2025milkywaystellarhalo}, its net angular momentum \citep{2017MNRAS.470.2959K,2017MNRAS.470.1259D,2018MNRAS.478..611B,2019ApJ...879..120C,2019MNRAS.486..378L,2019ApJ...871..184T,2020ApJ...899..110T,2021ApJ...919...66B}, and its rich chemo-dynamical diversity \citep{2021ApJ...919...66B,2024ApJ...974..167Z,yr1rrlyrae}. Some accretion events remain only partially mixed. Their debris can be observed as spatial overdensities \citep{2007ApJ...657L..89B,2008ApJ...673..864J,2009MNRAS.398.1757W,2014MNRAS.440..161S,2023ApJ...951...26C,2023ApJ...956..110C,2024arXiv240817250V,li2025milkywaystellarhalo} and stellar streams \citep[Jarvis et al. in preparation]{2023MNRAS.524.2124P,2025ApJ...985L..22N,2025ApJ...980...71V,2025MNRAS.539.2718P,2026arXiv260311171M} that continue to reveal the MW's ongoing accretion history \citep{1999Natur.402...53H,2006ApJ...642L.137B,2005ApJ...635..931B,2010MNRAS.406..744C}. 

Although the majority of these merger remnants are now well phase mixed into a seemingly smooth density background in the stellar halo, they continue to retain identifiable signatures in velocity and chemical abundance space \citep[Kizhuprakkat et al. in preparation]{Naidu2020,Naidu:2021aa,2025ApJ...985...47L}. For example, the debris of the Gaia-Sausage/Enceladus \citep[GSE;][]{2018MNRAS.478..611B,2018ApJ...863..113H,2018Natur.563...85H,2019MNRAS.486..378L,2021ApJ...923...92N,2022ApJ...924...23W,2023NatAs...7.1481H,2026ApJ...998..327D} merger is found to be confined to highly radial orbits and exhibits a significantly tighter metallicity dispersion than the surrounding halo population \citep{2018Natur.563...85H,Han_stellar_halo_density_profile,2024ApJ...974..167Z,yr1rrlyrae}. Consequently, precise 6-dimensional (6D) phase-space (positions, distances, and full space velocities) and high-resolution elemental-abundance measurements can help us better decode the merger history of the stellar halo. 

The launch of the {\it Gaia} space mission \citep{2016A&A...595A...1G} has revolutionized our view of the MW. By combining the precise proper motion measurements from {\it Gaia} and accurate distances, line-of-sight velocity, and abundance measurements from spectroscopic surveys, such as, SEGUE \citep{2009AJ....137.4377Y}, APOGEE \citep{2017AJ....154...94M}, GALAH \citep{2017MNRAS.465.3203M}, H3 \citep{2019ApJ...883..107C}, LAMOST \citep{2012RAA....12.1197C}, MagE \citep{2025ApJ...988..156C}, and DESI \citep{desiInstrument,desiScience,desi-collaboration22a,Spectro.Pipeline.Guy.2023,SurveyOps.Schlafly.2023,Corrector.Miller.2023,FiberSystem.Poppett.2024,DESI2023b.KP1.EDR,DESI2024.VII.KP7B,2025arXiv250314745D}, we are now able to trace the merger history of the MW in full 6D phase and abundance space. 

Assuming the stellar halo is in dynamical equilibrium, its observable features, including mean metallicity, mean velocity, velocity dispersion, and the velocity anisotropy, can be linked to the underlying gravitational potential. Precisely measuring the velocity moments and velocity anisotropy of the MW stellar halo is crucial for constraining the dark matter distribution around our MW, through, for example, the Jeans equation \citep[e.g.][]{2010ApJ...720L.108G,2018RAA....18..113Z,2022MNRAS.516..731B,2022MNRAS.511.5536R}. The readers can check \cite{2020SCPMA..6309801W} for a review.

Recent {\it Gaia}-era studies, however, have accumulated compelling evidence that the outer stellar halo ($\gtrsim$40-50~kpc) is not in dynamical equilibrium. Simulations and observations have revealed clear quantitative confirmation that the ongoing merger of the Large Magellanic Cloud (LMC) leaves pronounced, observable imprints on the outer stellar halo \citep{2007ApJ...668..949B,2013ApJ...764..161K,2019ApJ...884...51G,2021Natur.592..534C,2021MNRAS.506.2677E,2023ApJ...951...26C,2024MNRAS.527..437V,2024MNRAS.534.2694S,2024MNRAS.531.3524Y,2025MNRAS.544.2434S,2025MNRAS.544.1820Y,2025ApJ...988..156C,2025arXiv251004735B,2025MNRAS.542..560B,2026arXiv260308788B,2026MNRAS.545f2111B}. As the LMC accelerates towards north, the disk and inner halo, which have shorter dynamical time scale, move towards south, i.e, the reflex motion of the MW disk, imprinting an equal-and-opposite dipole pattern on the outer stellar halo radial velocity field with respect to the observer, that is most pronounced beyond $\sim$~40-50 kpc. 

Following these N-body simulations results, \cite{2025MNRAS.542..560B} further 
fit the velocity field of the MW outer stellar halo ($>$50~kpc) by reconstructing the velocity distribution with a large sample of blue horizontal branch (BHB) stars. The authors confirmed that the outer stellar halo undergoes a net inward contraction and that the reflex motion of the outer stellar halo is organized into a coherent dipole pattern, likely induced by the perturbation from the infall of the LMC.

In this work, we assemble a large sample of K giants from the second Data Release (DR2) of the DESI Stellar Surveys. By combining the proper motions from {\it Gaia} and the precise distance, line-of-sight velocity, and abundance measurements from DESI, we measure the mean metallicity, mean velocity, velocity dispersion, and the velocity anisotropy profiles as a function of Galactocentric distance by fitting the joint distribution with a Gaussian mixture model (GMM). 
We also investigate the non-equilibrium outer stellar halo by fitting the velocity distribution with a dipole velocity field model.

This paper is organized as follows: Section~\ref{sec:data} introduces the DESI MWS data and presents the selections of the halo K giant sample. Section~\ref{sec: model} illustrates the methodology adopted in this work. Section~\ref{sec: results} presents the results of this paper, including fitting the stellar halo with a GMM and the results of the non-equilibrium outer stellar halo. 
We discuss and compare our results with previous studies in Section~\ref{sec:discussion}, and conclude in Section~\ref{sec:concl}.

\section{Data}
\label{sec:data}

\subsection{The DESI Stellar Surveys}

The nomenclature of DESI Stellar Surveys and their relations are described in detail in the Appendix of \cite{dey2025backupprogramdarkenergy}. It consists of the Milky Way Survey \citep[MWS;][]{2023ApJ...947...37C}, the Milky Way Backup Program \citep[MWBP;][]{dey2025backupprogramdarkenergy}, the Science Validation \citep[SV;][]{2022AJ....164..207D,DESI2023b.KP1.EDR,2024AJ....167...62D} phase of observations, the DESI Andromeda Region Kinematic Survey \citep{2023ApJ...944....1D}, and the Milky Way stream and dwarf galaxy Surveys \citep[e.g.][]{2025ApJ...980...71V,2025ApJ...994..134D,2025ApJ...993..249Y,2025arXiv251204477Q}.

In our analysis, we use combined data\footnote{The data is observed from May 2021 to April 2024, and will be publicly released as DESI DR2, which jointly delivers $\sim 12$ million stellar spectra.} from the MWS, MWBP, and SV, to have the maximum number of stars available in the dataset to analyze the stellar halo. 
Briefly, the DESI MWS is performed at bright time together with DESI Bright Galaxies \citep{2023AJ....165..253H}, with flux limits of $16<r<19$~mag. The DESI MWBP is performed when the observing conditions are poor. MWBP targets are solely selected from {\it Gaia}, which reaches lower Galactic latitudes and covers a magnitude range of $11.2<G<19$~mag in {\it Gaia} $G$-band. SV is the pre-survey validation field, and is used to calibrate pipelines and selection functions. SV observations are confined to specific fields, resulting in a much higher completeness than the MWS and MWBP samples.
From the combined data, we select halo K giants for our analysis.

In our analysis, we assume that the DESI MWS selection functions do not affect the velocity field. As demonstrated in \cite{li2025milkywaystellarhalo}, the selection function only varies with sky coordinates. Stars are randomly selected for spectroscopy at a fixed sky position, implying no correlation between completeness fraction and stellar velocity.

\subsection{DESI Stellar Parameter Catalog}
In this work, we adopt the stellar parameters and line-of-sight velocity from the internal DESI MWS \textsc{rvj} pipeline, which is an upgraded version of the \textsc{rvs} pipeline\footnote{DESI provides three pipelines specialized for stellar spectra (\textsc{rvs}, \textsc{sp}, and \textsc{wd}). The \textsc{wd} pipeline is developed to process white dwarf spectra, and \textsc{sp} data were not available at the time of this work.} employed for the DESI EDR and DR1 VACs \citep{2024MNRAS.533.1012K,koposov2025desidatarelease1}. Relative to \textsc{rvs}, \textsc{rvj} exploits Korg-synthesised templates \citep{2023AJ....165...68W} and delivers element-specific abundance ratios such as [Mg/Fe] instead of a global $\alpha$-abundance.

Stellar distances are drawn from the DESI \texttt{SpecDis2} VAC (Li et al., in preparation), a revised edition of the DR1 \texttt{SpecDis} catalog \citep{li2025specdisvalueaddeddistance}, using the dust map of \cite{dustmap}. \texttt{SpecDis2} infers distance by mapping the combined spectroscopic information and photometric colours to absolute magnitude with a Neural Network (NN) and trains dwarfs and giants through distinct models. By training main-sequence and giant stars separately instead of mixing all stars together as in \texttt{SpecDis}, \texttt{SpecDis2} reduces the median distance error for giants from $\sim$~40~\% in \texttt{SpecDis} to $\sim$~12~\%. Here, the distance error is estimated from the error spectra and NN model uncertainties \citep{li2025specdisvalueaddeddistance}. We further validate the measurements using member stars in globular clusters (GCs) \citep{2021MNRAS.505.5957B} and dwarf galaxies \citep{2022ApJ...940..136P}. The distance uncertainty is about 18\% for member stars in GCs and dwarf galaxies.

\subsection{Selection of K giants}
\label{sec:Selection of K giants}
We first select K giants with effective temperature ($T_\mathrm{eff}$) of $4500~\mathrm{K}<T_\mathrm{eff} <5500~\mathrm{K}$ and $r$-band absolute magnitude of $-3<M_{r}<2$ after extinction correction. 
With this selection, we get a sample that contains $\sim$~370,000 stars. 
We further remove the contamination from thin/thick disks. We remove the thin/thick disk stars with $\mathrm{[Fe/H]}>-0.85\,\mathrm{dex}$\footnote{The typical uncertainty of $\mathrm{[Fe/H]}$ is about 0.05 dex in our sample.} \citep{li2025milkywaystellarhalo} and vertical distance to the disk plane smaller than 2~kpc \citep{2021ApJ...919...66B}. After this selection, we retain about 150,000 stars. 

We subsequently remove member stars belonging to known substructures. The member stars in GCs and dwarf galaxies are removed by cross-matching with \citet{2021MNRAS.505.5957B} and \citet{2022ApJ...940..136P}. 
We then remove the Sagittarius (Sgr) stream geometrically, which is massive and may bias our conclusions about the remaining parts of the stellar halo
\footnote{In Section~\ref{sec:single_component_results}, we also present the results obtained without removing the Sgr stream, in order to examine the potential bias that the Sgr stream may induce in the velocity distribution of the stellar halo.}.
We adopt the stream-oriented coordinate system from \citet{10.1093/mnras/stt1862}, and compute the perpendicular latitude $\tilde{B}$ and remove all stars in the footprint of $|\tilde{B}| < 15$\footnote{We have cross-matched our Sgr-stream-depleted sample with the catalog compiled by \cite{2021MNRAS.501.2279V}, which identifies Sgr member stars based on {\it Gaia} astrometric and photometric measurements. Only $\sim$~10 stars in our sample were found to have counterparts in this reference catalog, indicating that our geometric selection effectively eliminates the bulk of contamination from the Sgr stream.}. 
Other stellar streams and substructures, including Sequoia, the Helmi stream, Thamnos, and Arjuna as identified in \cite{Naidu2020}, are retained in our sample.
These criteria yield a clean halo K giant sample of $\sim$~94,000 stars.

Finally, we remove stars with large velocity uncertainties. 
We adopt the solar position $(X_\odot, Y_\odot, Z_\odot) = (-8.122, 0, 0.0208)~\mathrm{kpc}$ and velocity $(V_{X,\odot}, V_{Y,\odot}, V_{Z,\odot}) = (12.9, 245.6, 7.78)~\mathrm{km~s^{-1}}$.
We then derive the Galactocentric Cartesian coordinates ($X, Y, Z$) and the Galactocentric spherical velocities ($V_r$\footnote{Hereafter, we call $V_r$ radial velocity, which is Galactocentric, to be distinguished from the Heliocentric line-of-sight velocities.}, $V_\phi$, $V_\theta$)\footnote{Throughout this paper, we define $V_r<0$ as moving towards the Galactic center, and define $V_\phi<0$ as prograde azimuthal rotation sharing the same sign as the disc rotation. Moreover, positive polar velocity, $V_\theta$ is defined as pointing towards the north celestial pole, which shares the same definition as $v_b$ in \cite{2025MNRAS.544.2434S}.} for the stars using \texttt{ASTROPY} \citep{2022ApJ...935..167A}. 
We then propagate uncertainties in the distances, proper motions from {\it Gaia} \citep{2016A&A...595A...1G} DR3, and line-of-sight velocity via 1000 Monte Carlo sampling, and get the corresponding uncertainties for different velocity components ($\delta V_r, \delta V_\phi, \delta V_\theta$). 
We only retain the stars with $|\delta V_r|<100~\mathrm{km~s^{-1}}, \ |\delta V_\phi|<100~\mathrm{km~s^{-1}},$ and $|\delta V_\theta|<100~\mathrm{km~s^{-1}}$\footnote{After this selection,  typical uncertainty of $|\delta V_r|, \ |\delta V_\phi|,|\delta V_\theta|$ are about $7~\mathrm{km~s^{-1}}, 31~\mathrm{km~s^{-1}},$ and $|\delta V_\theta|<29~\mathrm{km~s^{-1}}$, respectively.}. 
After this selection, we get the final halo K giant sample, which contains $\sim$~64,000 stars and covers a Galactocentric distance ($r_\mathrm{GC}$) range from 3~kpc to 160~kpc.

\section{Method}
\label{sec: model}
In Section~\ref{sec:Bayesian Mixture Model}, we first introduce the GMM for characterizing the velocity field of different stellar populations.  We then illustrate how to characterize the anisotropic velocity distribution of the outer stellar halo ($50~\mathrm{kpc}<r_\mathrm{GC}<120~\mathrm{kpc}$) with a dipole velocity field in Section~\ref{sec:dipole_velocity_field_model}.

\subsection{Gaussian Mixture Model}
\label{sec:Bayesian Mixture Model}

Following \cite{yr1rrlyrae}, we assume that the stellar halo can be effectively described by a metal-rich (MR) component\footnote{In this work, the term component and population are used interchangeably.} and a metal-poor (MP) component. These two stellar populations can be distinguished in both metallicity and velocity space, and their joint distribution is well described by a GMM.

We first assume that the joint distributions of the MR or MP component
are multivariate Gaussians in $V_r$, $V_\phi$, $V_\theta$\footnote{We compute the covariance matrix of the three velocity components and find that the off-diagonal elements are at least an order of magnitude smaller than the diagonal elements, indicating that correlations between the velocity components are weak.
Hence, we believe that the assumption of approximately independent Gaussian velocity components is reasonable for the analysis below.
}, and [Fe/H] space:
\begin{equation}
\mathcal{L}_i^{\text{class}} = \mathcal{L}_{v_r,i}^{\text{class}} \times \mathcal{L}_{v_{\phi},i}^{\text{class}} \times \mathcal{L}_{v_{\theta},i}^{\text{class}} \times\mathcal{L}_{[\text{Fe/H}],i}^{\text{class}},
\end{equation}
Here, ``class'' stands for ``MR'' or ``MP'', $i$ stands for the $i$-th star, and a single Gaussian distribution is defined by
\begin{equation}
\mathcal{L}_{x,i}^{\text{class}} = \frac{1}{\sqrt{2\pi(\sigma_x^2+\delta_{x,i}^2)}} \exp\left[-\frac{(x_i-\langle x\rangle)^2}{2(\sigma_x^2+\delta_{x,i}^2)}\right].
\label{equ:single_guassian}
\end{equation}
Here, $x$ refers to  $V_r$, $V_\phi$, $V_\theta$, or [Fe/H]. $\langle x\rangle$, $\sigma_x$ are the mean and the standard deviation of the Gaussian distribution, and $\delta_x$ is the observational uncertainty. Throughout this paper, we denote the mean velocities and velocity dispersions in Galactocentric spherical coordinates as $\langle V_r\rangle$, $\langle V_\phi\rangle$, $\langle V_\theta\rangle$ and $\sigma_r$, $\sigma_\phi$, $\sigma_\theta$. We then assume that the fraction of the MR component is $p_{\text{MR}}$ ($0<p_{\text{MR}}<1$), and thus the final joint distribution or likelihood ($\mathcal{L}^{\text{tot}}$) of the stellar halo  can be defined as:
\begin{equation}
\ln(\mathcal{L}^{\text{tot}}) = \sum_{i=1}^{N} \ln[p_{\text{MR}} \, \mathcal{L}_i^{\text{MR}} + (1 - p_{\text{MR}}) \, \mathcal{L}_i^{\text{MP}}].
\label{equ:double_components_likelihood}
\end{equation}
There are a total of 17 free parameters in this model: eight for the means and standard deviations of the Gaussian distributions and one for the fraction of MR halos. Hereafter, we call this the double-component model.
The probability for the i-th star that belongs to the MR stellar halo component is:
\begin{equation}
P_i^\text{MR} = \frac{p_{\text{MR}} \mathcal{L}_i^{\text{MR}}}{p_{\text{MR}} \mathcal{L}_i^{\text{MR}} + (1 - p_{\text{MR}}) \mathcal{L}_i^{\text{MP}}}.
\label{equ:possibity_belongs_to_MR_halo}
\end{equation}
We further assume a uniform prior for all parameters\footnote{$\langle V_x\rangle^\text{class} \in [-100, 100] \, \mathrm{km~s^{-1}}$, $\sigma_x^\text{class} \in [10, 200] \, \mathrm{km~s^{-1}}$, $\mu_\text{[Fe/H]}^\text{class} \in [-3,-0.85] \, \text{dex}$, $\sigma_\text{[Fe/H]}^\text{class} \in [0.05,0.6] \, \text{dex}$, and $p_\text{MR} \in [0,1]$, 
where $x=r,\phi,\theta$ and $\text{class}=\text{MR},\text{MP}$. We have also tested a range of alternative prior settings and confirmed that our results are not sensitive to the prior choices.} in the double-component model and use Markov Chain Monte Carlo (MCMC) sampling to sample the posterior distribution with the Python package \texttt{emcee} \citep{emcee}. We employ 100 walkers, discarding the first 400 steps as before burn-in, and run 1000 steps in total. 

To compare results from the doule-component model, we also fit the velocity field with only one component (i.e. $p_{\text{MR}}=1$). Hereafter, we call it the single-component model. There are only 8 free parameters in this model (four means and four standard deviations). 

We present the best-fit results from the single-component and double-component models in Section~\ref{sec:single_component_results} and Section~\ref{sec:doubel_components_results1},~\ref{sec:doubel_components_results2}, respectively.

\subsection{Dipole Velocity Field Model}
\label{sec:dipole_velocity_field_model}
According to N-body simulations \citep{2019ApJ...884...51G,2024MNRAS.534.2694S,2024MNRAS.527..437V,2024MNRAS.531.3524Y,2025MNRAS.544.2434S,2025MNRAS.544.1820Y} or rigid MW-LMC simulations with stellar halo particles \citep{2026MNRAS.545f2111B}, the interaction between LMC and MW will strongly perturb the velocity field of the stellar halo in the outskirts ($r_\mathrm{GC}>50~\mathrm{kpc}$), manifesting as the density contrast \citep[e.g.][]{2021Natur.592..534C,joao,li2025milkywaystellarhalo}, and variable mean velocity and velocity dispersion at distinct line-of-sight directions \citep{2021NatAs...5..251P,2025ApJ...988..156C,2025MNRAS.542..560B}. 

As the LMC accelerates towards the north, the MW disk and inner halo move towards south, leading to a global positive radial velocity (away from the observer) observed in the northern outer halo and a global negative velocity (towards the observer) in the southern outer halo. 

Following \cite{2025MNRAS.542..560B}, we characterize the reflex motion for the outer stellar halo. We assume that the $V_r$ distribution obeys a Gaussian distribution with a position-dependent mean velocity $\mu_r$:
\begin{equation}
\mathcal{L}_{V_r,i} = \frac{1}{\sqrt{2\pi(\sigma_r^2+\delta_{r,i}^2)}} \exp\left[-\frac{(V_r-\mu_r)^2}{2(\sigma_r^2+\delta_{r,i}^2)}\right].
\label{equ:LMC_perturbation_likelihood}
\end{equation}
Here $\mu_\mathrm{r}$ is dependent on sky position:
\begin{equation}
\mu_\mathrm{r} = \mu_{\text{compr}} + \mu_{\text{dipole}} \cdot \cos(\theta),
\label{equ:LMC_mean_velocity_field}
\end{equation}
here the parameter $\mu_\mathrm{compr}$ quantifies the bulk radial motion of the outer stellar halo. The parameter $\mu_\mathrm{compr}$ adopted here quantifies the net inward contraction or outward expansion of the outer stellar halo, with a negative sign representing contraction. The parameter $\mu_{\text{dipole}}$ quantifies the amplitude of the dipole velocity field in the outer stellar halo, which reflects the reflex motion of the MW disk due to the infall of LMC (see Section~\ref{sec: lmc pertubation} for details).
$\theta$ is the angular distance between the position of a given star \( i \) in Galactic coordinates \( (l_i, b_i) \) and the direction of dipole apex \( (l_{\text{apex}}, b_{\text{apex}}) \). 
Hereafter, we call it the dipole velocity field model\footnote{Here, we assume a uniform prior for all parameters: $\mu_\text{compr} \in [-100,100] \, \mathrm{km~s^{-1}}$, $\mu_\text{dipole} \in [-100,100] \, \mathrm{km~s^{-1}}$, $\sigma_r \in [10,200] \, \mathrm{km~s^{-1}}$, $l_\text{apex} \in [0,2\pi]$, and $\cos{b_\text{apex}} \in [-1,1]$.}. We run MCMC to get the best-fit $V_r$ distribution of the outer stellar halo in Section~\ref{sec: lmc pertubation}.

\section{Results}
\label{sec: results}
In this section, we present the results of the best-fit single-component model in Section~\ref{sec:single_component_results}, and then show the double-component model in Section~\ref{sec:doubel_components_results1} and \ref{sec:doubel_components_results2}. Finally, we present the best-fit results of the dipole velocity field model in Section~\ref{sec: lmc pertubation}.

\begin{figure}
\begin{center}
\includegraphics[width=0.49\textwidth]{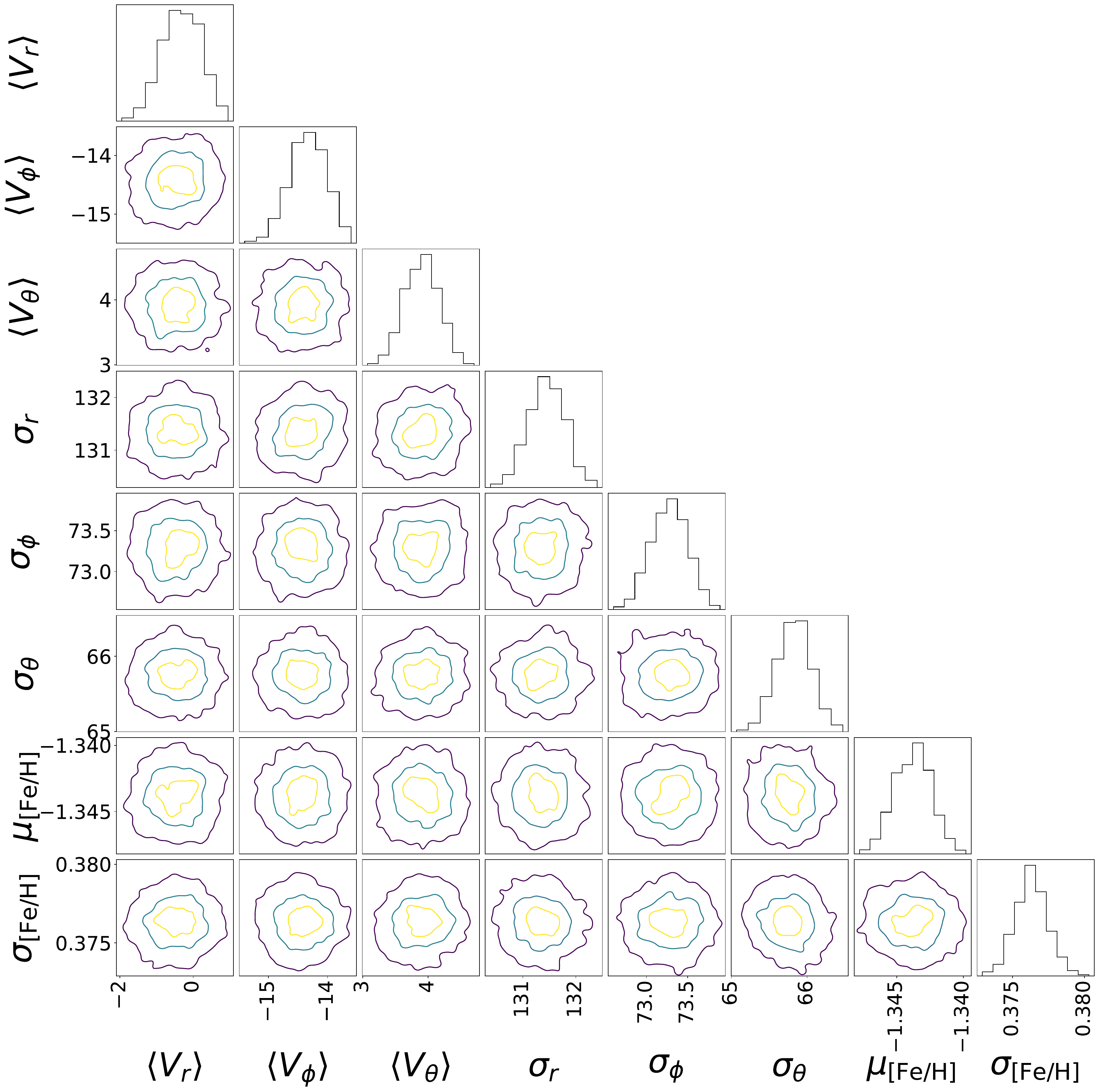}%
\end{center}
\caption{Error contours for different parameter combinations of the single-component model. Histograms show the marginalized posterior probability distributions for each parameter. $\langle V_r\rangle$, $\langle V_\phi\rangle$, and $\langle V_\theta\rangle$, which denote the three components of the mean velocity vector in Galactocentric spherical coordinates. $\sigma_r$, $\sigma_\phi$, and $\sigma_\theta$ are the corresponding velocity dispersions. $\mu_\mathrm{[Fe/H]}$ and $\sigma_\mathrm{[Fe/H]}$ characterize the mean and variance of the metallicity distribution, respectively.
The yellow, light blue, and dark blue contours represent the $30\%$, $1\sigma$, and $2\sigma$ regions of the MCMC post-burn distributions, respectively.}
\label{fig:mcmc_one_component}
\end{figure}

\subsection{Single-Component Model Results}
\label{sec:single_component_results}

\begin{figure*}
\begin{center}
\includegraphics[width=0.99\textwidth]{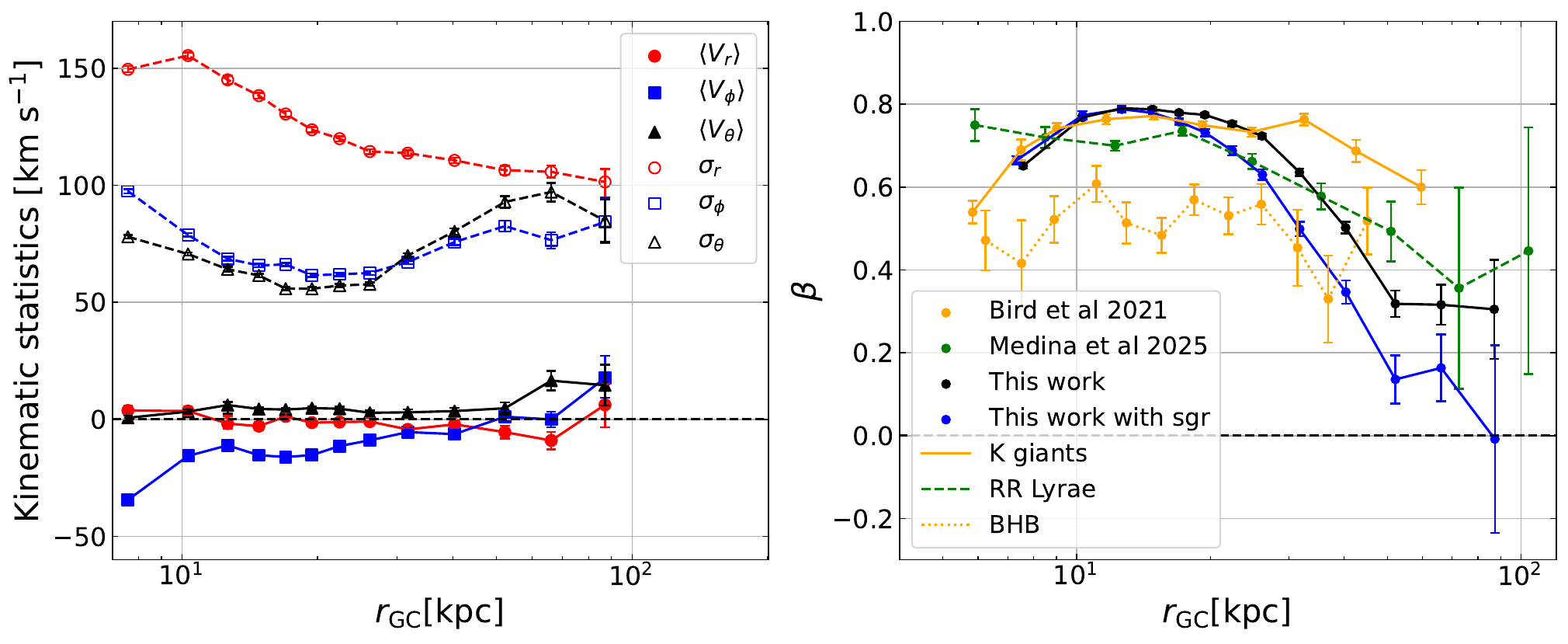}%
\end{center}
\caption{{\bf Left:} Distance ($r_\mathrm{GC}$) dependence of the mean velocity (filled markers connected with solid lines) and velocity dispersion (open markers connected with dashed lines). The error bars represent the 1$\sigma$ uncertainties.
{\bf Right:} Velocity anisotropy ($\beta$) as a function of distance. The 1$\sigma$ uncertainties of velocity anisotropy are propagated from the uncertainties in velocity dispersion ($\sigma_r, \sigma_\phi, \sigma_\theta$). The black/blue, orange, and green dots denote the best-fit model in this work with/without masking Sgr stream, \cite{2021ApJ...919...66B} and \cite{yr1rrlyrae}, respectively. 
Different line styles denote distinct tracer types: solid lines correspond to K giants, dashed line to RR Lyrae stars, and dotted line to BHB stars. 
The black horizontal dashed line denotes $\beta=0$, where $\beta>0$ corresponds to radial dominance and $\beta<0$ corresponds to tangentially dominance.}
The $x-$axes of two panels stand for the median value of distance at a given $r_\mathrm{GC}$ range.
\label{fig:single_component_result}
\end{figure*}

\begin{table*}[ht]
\centering
\caption{Best-fit mean velocity and velocity dispersions of the single-component model in different $r_\mathrm{GC}$ bins. The $+-$ values correspond to 1$\sigma$ uncertainties. The penultimate column provides the measured velocity anisotropy parameter, $\beta$. The last column provides the number of stars in each bin.}
\renewcommand{\arraystretch}{1.8}
\scriptsize      
\setlength{\tabcolsep}{3.5pt}  
\begin{tabular}{@{}lcccccccccccc@{}}
\toprule
$r_\mathrm{GC}\mathrm{[kpc]}$&$\langle V_r\rangle\mathrm{[km~s^{-1}]}$ & $\langle V_\phi\rangle\mathrm{[km~s^{-1}]}$ & $\langle V_\theta\rangle\mathrm{[km~s^{-1}]}$ & $\sigma_r\mathrm{[km~s^{-1}]}$ & $\sigma_\phi\mathrm{[km~s^{-1}]}$ & $\sigma_\theta\mathrm{[km~s^{-1}]}$ & $\mu_\mathrm{[Fe/H]}\mathrm{[dex]}$& $\sigma_\mathrm{[Fe/H]}\mathrm{[dex]}$ & $\beta$ & N\\
\midrule
$[3,160]$ & $-0.47^{+0.5}_{-0.5}$& $-14.4^{+0.3}_{-0.3}$& $3.9^{+0.3}_{-0.3}$& $131.3^{+0.4}_{-0.3}$& $73.3^{+0.2}_{-0.3}$& $65.8^{+0.2}_{-0.2}$& $-1.344^{+0.002}_{-0.001}$& $0.376^{+0.001}_{-0.001}$ & $0.719^{+0.003}_{-0.003}$ &  $64573$\\
$[3,9)$ & $3.8^{+1.8}_{-1.9}$& $-34.3^{+1.3}_{-1.2}$ & $0.68^{+1.0}_{-1.1}$ & $149.5^{+1.3}_{-1.4}$& $97.5^{+0.9}_{-0.9}$& $78.0^{+0.8}_{-0.9}$& $-1.335^{+0.005}_{-0.005}$& $0.395^{+0.004}_{-0.004}$& $0.651^{+0.012}_{-0.012}$& $6987$\\
$[9,11.5)$ & $3.56^{+1.8}_{-1.9}$& $-15.6^{+1.0}_{-1.1}$ & $3.3^{+0.9}_{-0.9}$ & $155.4^{+1.3}_{-1.3}$& $78.8^{+0.8}_{-0.8}$& $70.6^{+0.7}_{-0.8}$& $-1.372^{+0.005}_{-0.005}$& $0.404^{+0.003}_{-0.003}$& $0.768^{+0.008}_{-0.008}$ & $6890$\\
$[11.5,13.7)$ & $-1.7^{+2.1}_{-3.2}$& $-11.1^{+1.5}_{-1.2}$ & $6.1^{+1.2}_{-3.7}$ & $145.1^{+1.8}_{-1.7}$& $68.6^{+0.9}_{-0.9}$& $64.1^{+1.0}_{-0.9}$& $-1.365^{+0.009}_{-0.008}$& $0.396^{+0.004}_{-0.009}$& $0.790^{+0.011}_{-0.010}$ & $6741$\\
$[13.7,15.9)$ & $-2.8^{+1.7}_{-1.8}$& $-15.3^{+0.9}_{-1.1}$ & $4.4^{+0.9}_{-0.9}$ & $138.3^{+1.3}_{-1.2}$& $65.7^{+0.7}_{-0.7}$& $61.5^{+0.8}_{-0.7}$& $-1.341^{+0.006}_{-0.005}$& $0.366^{+0.004}_{-0.004}$& $0.788^{+0.009}_{-0.008}$ & $6671$\\
$[15.9,18.2)$ & $1.2^{+1.5}_{-1.6}$& $-16.1^{+0.9}_{-0.9}$ & $4.2^{+0.8}_{-0.8}$ & $130.5^{+1.2}_{-1.2}$& $66.2^{+0.7}_{-0.7}$& $55.8^{+0.6}_{-0.7}$& $-1.324^{+0.005}_{-0.005}$& $0.361^{+0.003}_{-0.003}$& $0.780^{+0.009}_{-0.008}$ & $6640$\\
$[18.2,20.8)$ & $-1.3^{+1.6}_{-1.5}$& $-15.2^{+1.0}_{-1.0}$ & $4.8^{+0.8}_{-0.8}$ & $123.7^{+1.1}_{-1.1}$& $61.5^{+0.7}_{-0.8}$& $55.8^{+0.7}_{-0.8}$& $-1.307^{+0.005}_{-0.005}$& $0.356^{+0.003}_{-0.004}$& $0.775^{+0.009}_{-0.010}$ & $6475$\\
$[20.8,24.2)$ & $-1.1^{+1.6}_{-1.6}$& $-11.3^{+1.0}_{-1.0}$ & $4.5^{+0.9}_{-0.8}$ & $119.9^{+1.2}_{-1.1}$& $62.0^{+0.8}_{-0.8}$& $57.1^{+0.7}_{-0.8}$& $-1.311^{+0.005}_{-0.005}$& $0.355^{+0.004}_{-0.004}$& $0.753^{+0.010}_{-0.010}$ & $6242$\\
$[24.2,28.7)$ & $-0.9^{+1.5}_{-1.5}$& $-8.9^{+0.9}_{-1.0}$ & $2.8^{+0.9}_{-0.9}$ & $114.5^{+1.1}_{-1.0}$& $62.5^{+0.8}_{-0.8}$& $57.7^{+0.7}_{-0.8}$& $-1.319^{+0.005}_{-0.005}$& $0.348^{+0.004}_{-0.004}$& $0.723^{+0.012}_{-0.012}$ & $5879$\\
$[28.7,35.8)$ & $-4.2^{+1.5}_{-1.5}$& $-5.4^{+1.1}_{-1.1}$ & $3.0^{+1.2}_{-1.2}$ & $113.8^{+1.1}_{-1.1}$& $67.2^{+0.9}_{-0.9}$& $70.0^{+1.0}_{-0.9}$& $-1.359^{+0.005}_{-0.006}$& $0.373^{+0.004}_{-0.004}$& $0.636^{+0.018}_{-0.018}$ & $5318$\\
$[35.8,47.5)$ & $-2.1^{+1.6}_{-1.7}$& $-6.2^{+1.4}_{-1.3}$ & $3.5^{+1.4}_{-1.4}$ & $110.7^{+1.2}_{-1.2}$& $75.8^{+1.3}_{-1.3}$ & $80.3^{+1.4}_{-1.3}$& $-1.386^{+0.006}_{-0.006}$& $0.377^{+0.005}_{-0.005}$& $0.502^{+0.029}_{-0.028}$ & $4190$\\
$[47.5,60)$ & $-5.4^{+2.7}_{-2.7}$& $1.1^{+2.4}_{-2.4}$ & $4.7^{+2.5}_{-2.8}$ & $106.4^{+1.9}_{-1.9}$& $82.6^{+2.4}_{-2.3}$& $92.9^{+2.5}_{-2.3}$& $-1.422^{+0.011}_{-0.010}$& $0.392^{+0.008}_{-0.008}$& $0.318^{+0.063}_{-0.062}$ & $1596$\\
$[60,80)$ & $-8.9^{+3.8}_{-3.5}$& $0.1^{+3.4}_{-3.3}$ & $16.5^{+4.2}_{-4.1}$ & $105.7^{+2.6}_{-2.7}$& $76.5^{+3.6}_{-3.4}$& $97.2^{+4.1}_{-3.4}$& $-1.415^{+0.015}_{-0.014}$& $0.380^{+0.011}_{-0.011}$& $0.316^{+0.096}_{-0.094}$ & $779$\\
$[80,160]$ & $6.2^{+9.5}_{-8.8}$& $17.6^{+8.4}_{-9.5}$ & $14.6^{+8.8}_{-8.7}$ & $101.4^{+6.6}_{-5.5}$& $84.3^{+9.1}_{-10.6}$& $84.8^{+9.0}_{-10.6}$& $-1.420^{+0.042}_{-0.031}$& $0.372^{+0.026}_{-0.026}$& $0.305^{+0.239}_{-0.226}$ & $164$\\
\bottomrule
\end{tabular}
\label{table:single_component_results}
\end{table*}

We first present the results of the single-component model (i.e. $P_\mathrm{MR}=1$, see Equation~\ref{equ:double_components_likelihood}). 
We model the global velocity and metallicity distribution of the stellar halo using all halo stars that span $3~\mathrm{kpc}<r_\mathrm{GC}<160~\mathrm{kpc}$, and subsequently repeat the fitting process within discrete radial bins to resolve distance-dependent variations. 

Figure~\ref{fig:mcmc_one_component} shows the error contours for different parameter combinations and the marginalized posterior distribution for each parameter, using all halo K giants ($3~\mathrm{kpc}<r_\mathrm{GC}<160~\mathrm{kpc}$). The best-fit parameters and $1\sigma$ uncertainties are presented in the first row of Table~\ref{table:single_component_results}.
For our best-fit model, we find that: (1) The stellar halo exhibits an average velocity anisotropy $\beta$ of $\sim0.72$, which demonstrates that the orbits of K giants are strongly radial\footnote{Here, we define the velocity anisotropy as $\beta=1-\frac{\sigma_\phi^2+\sigma_\theta^2}{2\sigma_r^2}$. It characterizes the orbital geometry of stars: $\beta>0$ indicates radially dominated orbits, $\beta=0$ denotes isotropic orbits, and $\beta<0$ corresponds to tangentially dominated orbits.}.
(2) The stellar halo exhibits a tiny net rotation, reflected by a mean azimuthal velocity of $\langle V_\phi\rangle=-14.4~\mathrm{km~s^{-1}}$. Here, the negative sign indicates that the net azimuthal motion of the stellar halo is in the same direction as the rotation of the Galactic disk. The rotation signal exhibits a comparable amplitude to previous studies \citep{2017MNRAS.470.1259D,2019ApJ...871..184T,2020ApJ...899..110T,2021ApJ...919...66B}. While the mean radial velocity and polar velocity are nearly zero.

We further investigate the features of the kinematic velocity field across various radii. Table~\ref{table:single_component_results} summarizes the best-fit parameters and $1\sigma$, together with the associated $r_\mathrm{GC}$ ranges. 
The best-fit mean velocity, velocity dispersions, and their respective 1$\sigma$ uncertainties are derived from the posterior distributions of MCMC, and we avoid repeatedly showing the posterior contours of these models. 

\subsubsection{Radial dependent mean velocity and velocity dispersion}

In Figure~\ref{fig:single_component_result}, each symbol corresponds to a measurement in a radially-binned subsample, whose Galactocentric distance range is provided in the first column of Table~\ref{table:single_component_results}, and the $x$-axis for each symbol denotes the median Galactocentric distance of the stars within the corresponding radial interval.

The left panel of Figure~\ref{fig:single_component_result} displays the mean velocity and velocity dispersion profiles as a function of Galactocentric distance. 
Across the radial interval $3~\mathrm{kpc}<r_\mathrm{GC}<50~\mathrm{kpc}$, both the mean radial velocity $\langle V_r\rangle$ and mean polar velocity $\langle V_\theta\rangle$ remain statistically consistent with zero, consistent with the overall fitting of the stellar halo (Figure~\ref{fig:mcmc_one_component}). 
For the outer stellar halo ($r_\mathrm{GC}>50~\mathrm{kpc}$), the mean radial velocity is slightly biased from zero, indicating an inward motion of the outer stellar halo due to the perturbation from the LMC, and we will discuss this point in Section~\ref{sec: lmc pertubation} below. 

The outer stellar halo also shows a net positive motion in the polar direction, with an amplitude of approximately 15$~\mathrm{km~s^{-1}}$, consistent with the observations\footnote{\cite{2024MNRAS.534.2694S} finds a positive polar velocity in the outer stellar halo ($r_\text{GC}>50$kpc). \cite{2025ApJ...988..156C} also reports a $\langle V_\theta\rangle\sim15 \,\mathrm{km~s^{-1}}$ at $\sim80$ kpc (see Figure~5 in \cite{2025ApJ...988..156C}).} in \cite{2024MNRAS.534.2694S}, \cite{2025ApJ...988..156C} and the predictions by N-body simulations\footnote{Specifically, \cite{2025MNRAS.544.2434S} simulated the first passage of the LMC and demonstrated that the outer stellar halo exhibits a net positive polar velocity (see their Figure~4).} presented in \cite{2019ApJ...884...51G} and \cite{2025MNRAS.544.2434S}.
This is because the inner halo with a shorter dynamical time scale is moving towards the south when the LMC is currently moving towards the north. On the other hand, the outer halo with a longer dynamical time remains more intact. The outer halo then shows a motion towards the north with respect to the observer, who is currently moving south.

\cite{2025MNRAS.544.2434S} further claims that LMC minimally affects the azimuthal velocity distribution of the MW stellar halo in their simulations because the orbital plane of the LMC is nearly perpendicular to the Galactic disk plane and does not impact the angular momentum of stars in the azimuthal velocity direction. Hence, the non-zero $\langle V_\phi\rangle$ might orgin from other merger events, rather than perturbations from the LMC. 

As presented in the left of Figure~\ref{fig:single_component_result}, the mean azimuthal velocity $\langle V_\phi\rangle$ exhibits negative values within 30~kpc, indicating some tiny prograde rotations, decreases to zero ($r_\mathrm{GC}\sim70~\mathrm{kpc}$), and subsequently presents a positive net rotation ($\sim$~15$~\mathrm{km~s^{-1}}$, retrograde rotation) in the outermost bin ($80~\mathrm{kpc}<r_\mathrm{GC}<160~\mathrm{kpc}$).
Interestingly, \cite{2025ApJ...988..156C} also reports a distance-dependent $\langle V_\phi\rangle$ profile (see Figure~11 in their paper).  $\langle V_\phi\rangle$ first decreases to zero ($r_\mathrm{GC}\sim80~\mathrm{kpc}$), and subsequently increases again with distance ($r_\mathrm{GC}>80~\mathrm{kpc}$). They show a retrograde motion with an amplitude of $\sim15~\mathrm{km~s^{-1}}$ at $r_\mathrm{GC}\sim120~\mathrm{kpc}$, consistent with the measurements in this work. However, the origin of a retrograde motion in the outer stellar halo is less clear; it might originate from other merger events.

In particular, the innermost halo region ($3~\mathrm{kpc}<r_\mathrm{GC}<9~\mathrm{kpc}$) exhibits the strongest net azimuthal rotation of $V_\phi\sim-34~\mathrm{km~s^{-1}}$, which may just reflect the residual contaminations from the thick disk. 
Although the thick disk does not extend to $R>12~\mathrm{kpc}$\footnote{Here, $R$ is the radius in cylindrical coordinates.} \citep{2024MNRAS.531.1730T}. At $r_\mathrm{GC}>9~\mathrm{kpc}$, the stellar halo could still be contaminated by some disk-associated substructures.
For example, there might be contamination from the Monoceros Ring, Anticenter stream \citep{2020MNRAS.492L..61L,2021A&A...646A..99R,2024ApJ...961...65Q,2026arXiv260114562L}, or the Aleph structure \citep{2020ApJ...901...48N}. These substructures can extend to $ Z\sim10~\mathrm{kpc}, R\sim20~\mathrm{kpc}$, which could survive from our selection criteria on vertical distance ($|z|>2~\mathrm{kpc}$, see Section~\ref{sec:Selection of K giants}).
These stars rotate around the Galactic center with azimuthal velocity ($V_\phi\sim-200~\mathrm{km~s^{-1}}$), consistent with the disk \citep{2020ApJ...901...48N}. Hence they could lead to a net azimuthal rotation in our fitting. 
However, it was shown by \cite{2024ApJ...961...65Q} that the Monoceros Ring and Anticenter stream are chemically consistent with the thin disk. Thus, our chemical selection should have eliminated most of the contamination from these structures.

In Section~\ref{sec:doubel_components_results2} below, we will show that even for the MP population, the stellar halo still exhibits net rotation in $V_\phi$ space, which will more robustly confirm the existence of non-zero angular momentum in MP halo stars. And the MR halo stars exhibit a nearly constant velocity anisotropy ($\beta\sim0.90$) at $r_\mathrm{GC}>9~\mathrm{kpc}$, indicating that any residual disk contamination is negligible beyond 9~kpc.
We leave discussions on possible mechanisms that induced a rotating inner stellar halo in Section~\ref{sec:discussion}.

The velocity dispersion trend depicted in the left panel of Figure~\ref{fig:single_component_result} reveals a more intricate pattern: 
the radial velocity dispersion ($\sigma_r$) reaches its peak at $\sim10~\mathrm{kpc}$ and then decreases with distance, while the azimuthal and polar velocity dispersion are first decreasing ($r_\mathrm{GC}<30~\mathrm{kpc}$) and then increasing ($r_\mathrm{GC}>30~\mathrm{kpc}$) with distance, reaching comparable amplitude with $\sigma_r$ at $r_\mathrm{GC}>60~\mathrm{kpc}$, indicating a more isotropic stellar halo beyond 60~kpc. The drop of radial velocity dispersion at the innermost stellar halo within 10~kpc may also be due to thick disk contamination, as the disk stars exhibit systematically lower radial velocity dispersions. 

\subsubsection{Radial dependent velocity anisotropy}

The right panel of Figure~\ref{fig:single_component_result} presents the velocity anisotropy profile measured in this work (black dots) as a function of distance, compared with two previous measurements of \cite{2021ApJ...919...66B} and \cite{yr1rrlyrae}. The orange dots \citep{2021ApJ...919...66B} are based on a K giant sample (solid line) and a BHB sample (dotted line) after removing substructures, and the green dots represent the measurements from the DESI Year-1 RR Lyrae catalog \citep{yr1rrlyrae}. 

With the single-component model, we find that the velocity anisotropy (black dots) measured in this work is increasing at $r_\mathrm{GC}<10~\mathrm{kpc}$, and remains almost constant ($\beta\sim0.77$) for $10~\mathrm{kpc}<r_\mathrm{GC}<30~\mathrm{kpc}$. It then decreases quickly beyond 30~kpc and becomes less radial ($\beta<0.5$) at $r_\mathrm{GC}>50~\mathrm{kpc}$. 
Here, we also point out that the velocity anisotropy of the innermost stellar halo could be underestimated due to the thick disk contamination.
We also present the velocity anisotropy measurements (blue dots) without masking the Sgr stream. This profile exhibits a trend similar to that of the previous one, but yields globally underestimated $\beta$ values at $r_\mathrm{GC}>20~\mathrm{kpc}$, consistent with the fact that the motion of the Sgr stream is tangentially dominated, hence it is necessary to remove the Sgr stream to avoid biasing the velocity anisotropy measurements.

\begin{figure*}
\begin{center}
\includegraphics[width=0.99\textwidth]{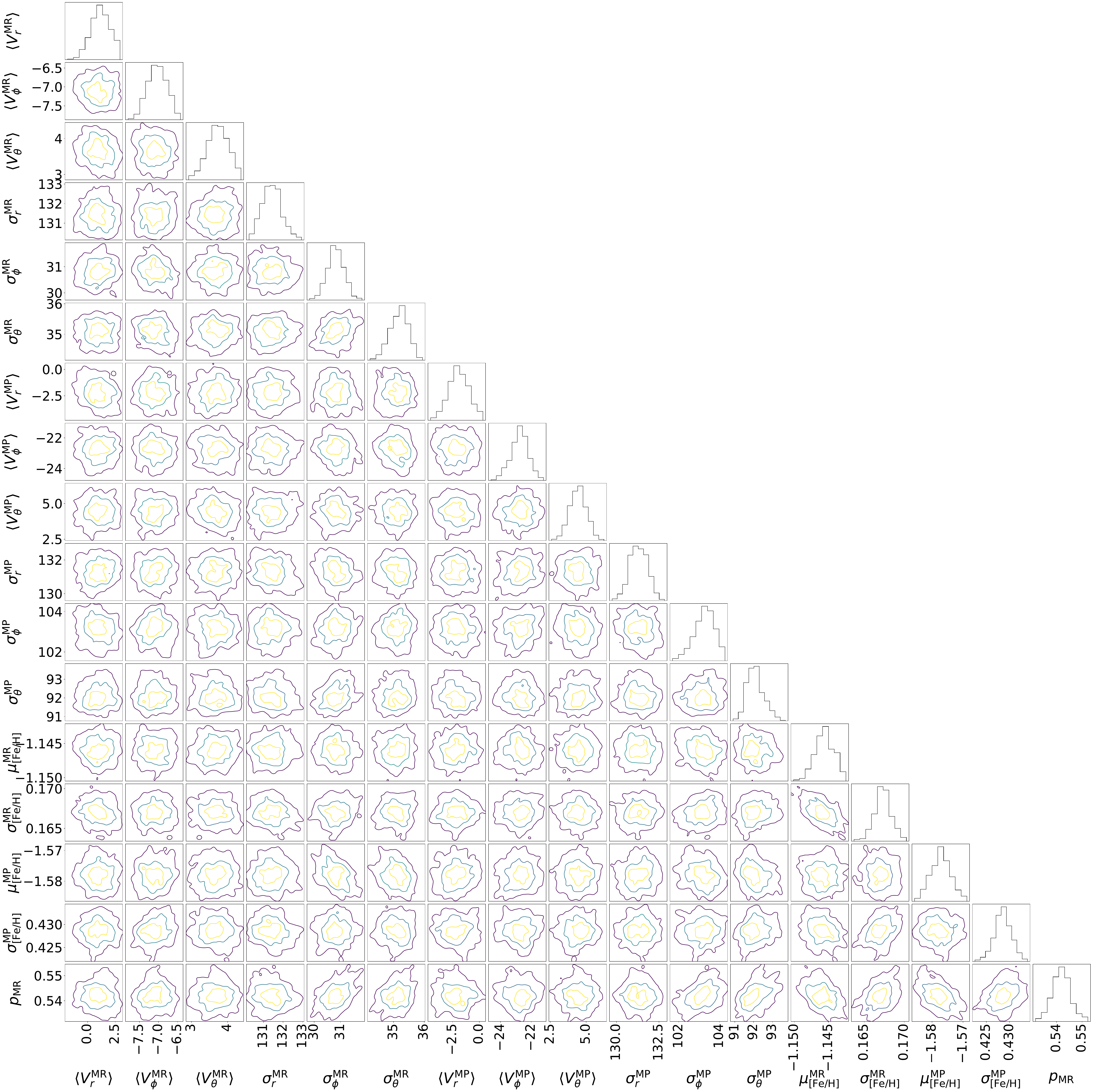}%
\end{center}
\caption{Error contours for different parameter combinations of the double-component model (see Equation~\ref{equ:double_components_likelihood}). Histograms show the marginalized posterior probability distributions for each parameter. $\langle V_x^\mathrm{MR}\rangle$ and $\langle V_x^\mathrm{MP}\rangle$ ($x=r,\phi,\theta$) are the means of the MR and MP halo populations in Galactocentric spherical coordinates. $\sigma_x^\mathrm{MR}$ and $\sigma_x^\mathrm{MP}$ ($x=r,\phi,\theta$) correspond to the velocity dispersions for two populations. $\mu_\mathrm{[Fe/H]}^\mathrm{MR/MP}$ and $\sigma_\mathrm{[Fe/H]}^\mathrm{MR/MP}$ are the mean and standard deviation of the metallicity distribution for two populations. $p_\mathrm{MR}$ is the fraction of the MR population in the stellar halo. The yellow, light blue, and dark blue contours represent the $30\%$, $1\sigma$, and $2\sigma$ regions of the MCMC post-burn distributions, respectively.}
\label{fig:mcmc_2components}
\end{figure*}

When comparing our $\beta$ measurements with those reported in \cite{2021ApJ...919...66B}, which were derived from a K giant sample, we find good consistency at $r_\mathrm{GC}<30~\mathrm{kpc}$.
Both datasets exhibit an increasing trend in the very inner stellar halo, followed by a nearly constant profile for the range $10~\mathrm{kpc}<r_\mathrm{GC}<30~\mathrm{kpc}$. However, in \cite{2021ApJ...919...66B} the stellar halo is still very radial ($\beta>0.5$) beyond 30~kpc. 
The $\beta$ measurements in \cite{yr1rrlyrae} exhibit a different behaviour: the velocity anisotropy is roughly monotonically decreasing. The velocity anisotropy is less radial for $10~\mathrm{kpc}<r_\mathrm{GC}<30~\mathrm{kpc}$ than our measurement, and becomes comparable to ours beyond 30~kpc. 
Finally, the $\beta$ measurements from \cite{2021ApJ...919...66B}, derived from a BHB giant sample, do not exhibit a significant dependence on Galactocentric distance; instead, they oscillate around a mean value of $\sim$~0.5.
The distance dependence of $\beta$ and the variation in $\beta$ values measured by different tracers may arise from intrinsic differences between the MR and MP populations, as well as variations in the mixing ratio of MR and MP populations across different samples. We present a detailed discussion in Section~\ref{sec:doubel_components_results2}.

Interestingly, due to the GSE merger event, our MW inner stellar halo is highly radial. This is in contrast to the expected average behaviour from galaxies in numerical simulations \citep[e.g.][]{He2024}, though the scatter of the measured velocity anisotropy profiles for simulated galaxies is huge.

Table~\ref{table:single_component_results} also provides the mean ($\mu_\mathrm{[Fe/H]}$) and standard deviation ($\sigma_\mathrm{[Fe/H]}$) of the metallicity distribution. But the $\mu_\mathrm{[Fe/H]}$ and $\sigma_\mathrm{[Fe/H]}$ do not show any obvious dependence on Galactocentric distance.

\subsection{Double-Component Model Results: metallicity distribution and metal-rich population fraction}
\label{sec:doubel_components_results1}

We now present the results of the double-component model. Consistent with Section~\ref{sec:single_component_results}, we first introduce the modeling results using the full halo star sample ($3~\mathrm{kpc}<r_\mathrm{GC}<160~\mathrm{kpc}$) after excluding the Sgr stream\footnote{We have also tried triple-component modeling by including the Sgr stream. In this case, we can well separate the Sgr stream, and the measurements for the other MR and MP components are largely similar to the results presented in this subsection. We thus do not repeatedly show the results for triple-component modeling.}. We then perform fits to the metallicity distribution and the fraction of the MR population across different distance bins. Fits to the velocity field will be presented in Section~\ref{sec:doubel_components_results2} below.

Figure~\ref{fig:mcmc_2components} shows the error contours for different parameter combinations and the marginalized posterior distribution for each parameter in the double-component model, using all halo K giants. The best-fit parameters and associated 1$\sigma$ uncertainties can be found in Table~\ref{table:double_component_results}.
For our best-fit model, we find that: 

(1) The fraction of the MR population $p_\mathrm{MR}$ is about 0.54, indicating that the MR and MP populations roughly have equal numbers of stars. 
However, as also demonstrated in \cite{li2025milkywaystellarhalo}, our metallicity-based selection of halo K giants (with a cut at $\mathrm{[Fe/H]}=-0.85\,\mathrm{dex}$)  excludes some MR GSE stars, which have $\mathrm{[Fe/H]}>-0.85\,\mathrm{dex}$. We emphasize that adopting a different metallicity cut may result in slightly different values of $p_\mathrm{MR}$.

(2) The MR population displays a markedly narrower metallicity dispersion than its MP counterpart: $\sigma_\mathrm{[Fe/H]}^\mathrm{MR}\sim0.17~\mathrm{dex}$ versus $\sigma_\mathrm{[Fe/H]}^\mathrm{MP}\sim0.43~\mathrm{dex}$. This agrees with the findings of \citet{yr1rrlyrae}, in which a compact distribution for the MR component 
($\sigma_\mathrm{[Fe/H]}^\mathrm{MR}\sim0.18~\mathrm{dex}$) and a substantially broader dispersion for the MP stellar halo ($\sigma_\mathrm{[Fe/H]}^\mathrm{MP}\sim0.47~\mathrm{dex}$) is reported. The distinct metallicity dispersions of the two populations may reflect different merger histories. In Section~\ref{sec:doubel_components_results2}, we further investigate the velocity anisotropy of the two components and suggest that the MR component is likely dominated by a single major merger event, whereas the MP population originates from multiple minor mergers, as indicated by their distinct dynamical properties.

(3) The MR population is confined to highly radial orbits, exhibiting a velocity anisotropy of $\beta_\mathrm{MR}\sim0.94$, whereas the MP population displays significantly less radial motion, with $\beta_\mathrm{MR}\sim0.46$. 

(4) Both populations exhibit a net azimuthal motion or slight rotation. However, the amplitude of $\langle V_\phi\rangle$ in the MP component is three times larger than that measured for MR halo stars ($\langle V_\phi^\mathrm{MR}\rangle\sim-7~\mathrm{km~s^{-1}}$ and $\langle V_\phi^\mathrm{MP}\rangle\sim-22~\mathrm{km~s^{-1}}$). This enhanced rotation signal suggests that the modest spin of the stellar halo is an intrinsic property rather than residual contamination from the Galactic disk or its associated structures, as the mean metallicity of the MP population is much poorer than that of the disk stars.

\begin{figure}
\begin{center}
\includegraphics[width=0.49\textwidth]{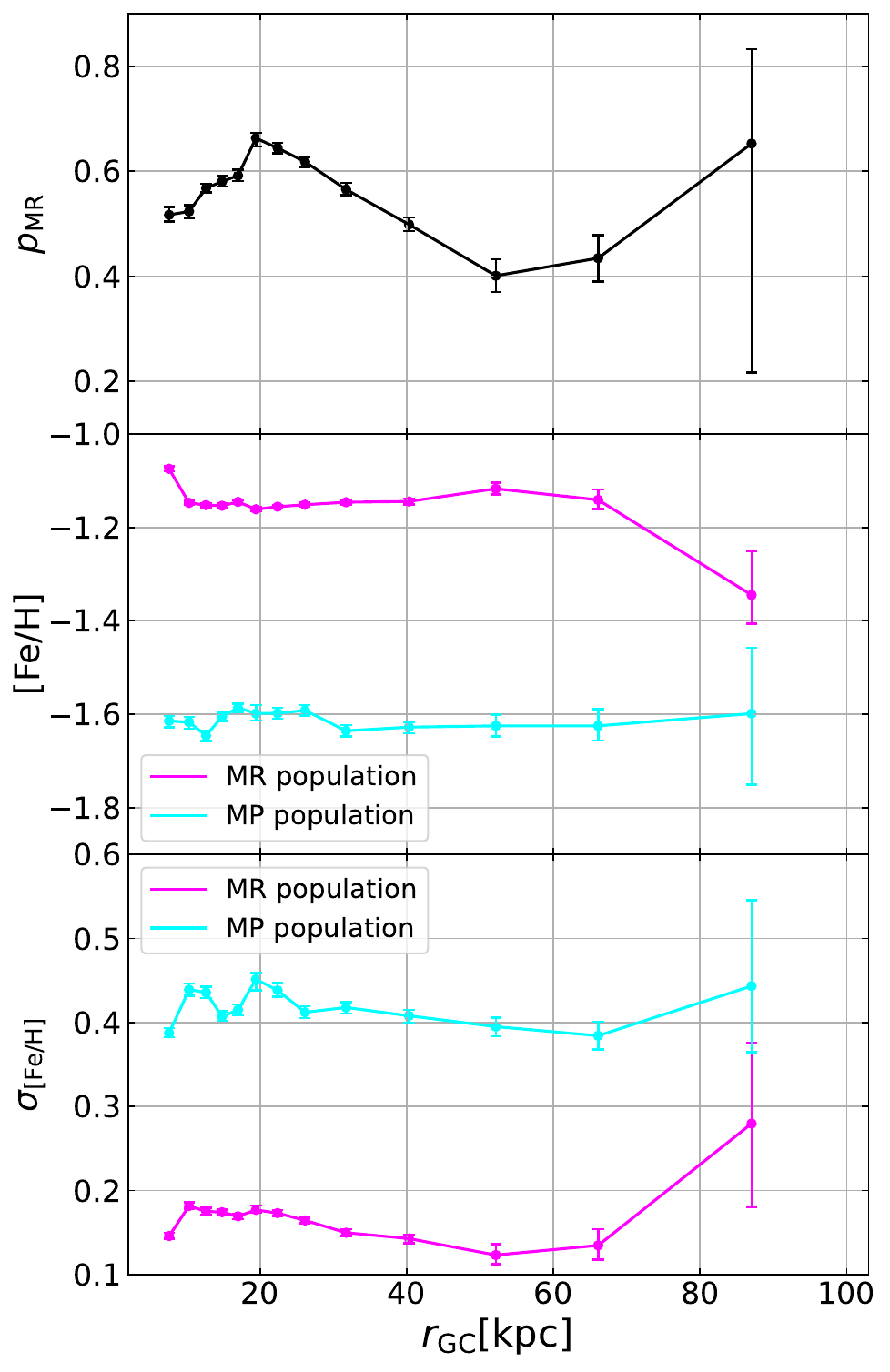}%
\end{center}
\caption{{\bf Top:} Fraction of the MR K giants as a function of Galactocentric distance $r_\mathrm{GC}$. {\bf Middle:} Mean metallicity profiles of the MR population (magenta dots) and the MP population (cyan dots). 
 {\bf Bottom:} Metallicity dispersion profiles of the MR population (magenta dots) and the MP population (cyan dots). 
The error bars stand for 1$\sigma$ uncertainties.}
\label{fig:fraction_of_metal_rich}
\end{figure}

We then proceed to dissect the stellar halo into multiple radial intervals and employ the double-component model to quantify the fraction of MR population, mean metallicity distribution, mean velocity, velocity dispersion, and velocity anisotropy of each population, as a function of Galactocentric distance.  
Table~\ref{table:double_component_results} compiles the best-fit parameters together with their $1\sigma$ uncertainties for every radial bin.


The top panel of Figure~\ref{fig:fraction_of_metal_rich} illustrates the fraction of the MR population ($p_\mathrm{MR}$) as a function of Galactocentric distance $r_\mathrm{GC}$. $p_\mathrm{MR}$ initially increases with Galactocentric distance ($r_\mathrm{GC}<20~\mathrm{kpc}$), subsequently experiences a rapid decline over $20~\mathrm{kpc}<r_\mathrm{GC}<50~\mathrm{kpc}$. Beyond 50~kpc, $\beta$ shows some rising trend, but the errorbars are significantly larger there.
As we will illustrate later in this section, the MR population is primarily contributed by GSE debris. The fraction of the MR population thus reflects the radial dependence of GSE star fraction.
However, we emphasize that our metallicity-based selection criterion may lead to a slight underestimation of the GSE star fraction.

Interestingly, \cite{yr1rrlyrae} also found a similar trend as the top panel of Figure~\ref{fig:fraction_of_metal_rich}. The authors calculated the possibility of a star that belongs to the MR stellar halo component ($P_i^\mathrm{MR}$, see Equation~\ref{equ:possibity_belongs_to_MR_halo}) and found that the fraction of stars with $P_i^\mathrm{MR}>0.5$ firstly increases at $r_\mathrm{GC}<20~\mathrm{kpc}$, and then undergoes a sharp decrease in the interval of $20~\mathrm{kpc}<r_\mathrm{GC}<50~\mathrm{kpc}$ (see Figure~16 in \cite{yr1rrlyrae}). The large extent of the MR stellar halo is also supported by \cite{2023ApJ...951...26C}, who reported that the GSE, which is more MR than other stellar halo components, could extend beyond $\sim$~50~kpc in toward some overdensity region directions, such as the Hercules-Aquila Cloud South/North overdensity \citep[HAC-N/-S;][]{2007ApJ...657L..89B,2009MNRAS.398.1757W,2014MNRAS.440..161S}, the Pisces overdensity \citep{2023ApJ...956..110C,2024arXiv240817250V} and the Virgo overdensity \citep[VOD;][]{2023ApJ...951...26C} directions.

However, the extended MR stellar halo in our measurement is at least partially affected by distance uncertainties, as the stars in the inner stellar halo will be scattered outwards due to Eddington bias. We provide detailed tests in Appendix~\ref{app:error_on_model} to quantify the effect, and it is shown that though the distance uncertainties indeed increase the MR fraction, the fractional change is not large enough to violate our conclusions.

The middle panel of Figure~\ref{fig:fraction_of_metal_rich} illustrates the metallicity profiles of the two populations. The mean metallicity of the MR component declines only within two radial ranges: $r_\mathrm{GC}<10~\mathrm{kpc}$  (reflecting contaminations from the thick disk), and $r_\mathrm{GC}>80~\mathrm{kpc}$. 
The mean metallicity of the MP population exhibits no significant dependence on Galactocentric distance. 
The nearly flat metallicity profile indicates that the MR and MP populations in our MW stellar halo are both roughly homogeneous. 

The bottom panel of Figure~\ref{fig:fraction_of_metal_rich}  further presents metallicity dispersions of both populations at various radii. The MP population shows a significantly broader metallicity distribution than the MR population, except for the outermost point at $\sim$~90~kpc. In this outermost region, $P_\mathrm{MR}$ has the greatest uncertainty, and the MR population has a mean metallicity more similar to the MP population, indicating the outermost halo may only require the MP population for a valid description.

\begin{figure*}
\begin{center}
\includegraphics[width=0.99\textwidth]{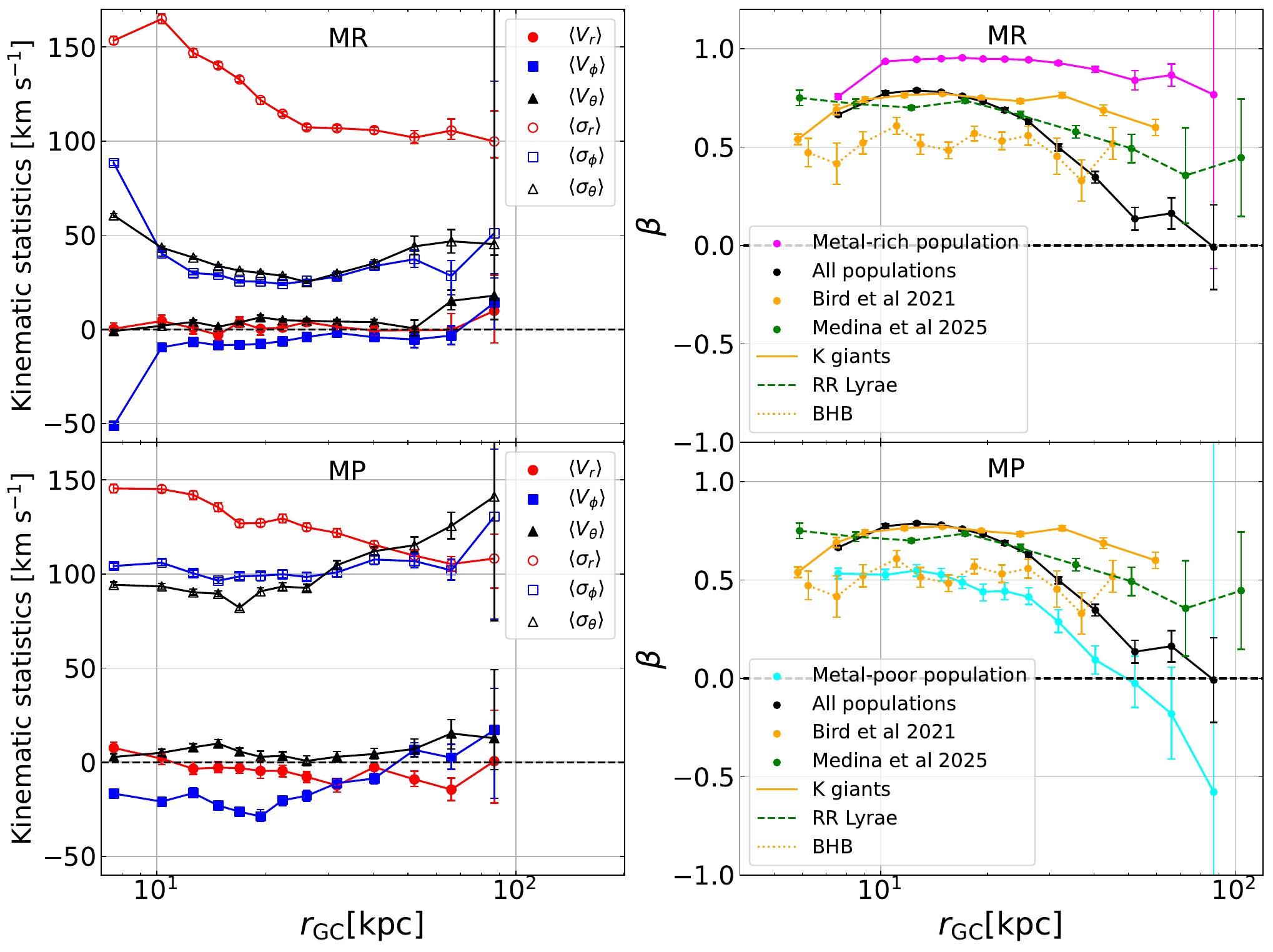}%
\end{center}
\caption{Distance dependence of velocity field for the MR population (first row) and the MP population (second row). In the left panel, the filled markers denote the mean velocity profile binned as a function of Galactocentric distance, and the open markers represent the velocity dispersion dependence on distance. The right panel shows the velocity anisotropy at different radii. The magenta/cyan dots stand for the MR/MP populations, and the black dots are from the best-fit single-component model in Section~\ref{sec:single_component_results} (see the right panel of Figure~\ref{fig:single_component_result}). 
The orange and green dots denote the best-fit model in \cite{2021ApJ...919...66B} and \cite{yr1rrlyrae}, respectively. 
Different line styles denote distinct tracer types: solid lines correspond to K giants, dashed line to RR Lyrae stars, and dotted line to BHB stars.
The error bars stand for 1$\sigma$ uncertainties. The black horizontal dashed line denotes $\beta=0$, where $\beta>0$ corresponds to radial dominance and $\beta<0$ corresponds to tangentially dominance.}
\label{fig:double_component_results}
\end{figure*}

\subsection{Double-Component Model Results: velocity field and spin}
\label{sec:doubel_components_results2}

\subsubsection{Radial dependent mean velocity and velocity dispersion of metal-rich and metal-poor components}

Figure~\ref{fig:double_component_results} shows the mean velocity, velocity dispersion, and velocity anisotropy as a function of Galactocentric distance. 
The best-fit parameters and associated 1$\sigma$ uncertainties can be found in Table~\ref{table:double_component_results}.
The first and second rows of the left panel in Figure~\ref{fig:double_component_results} correspond to the MR and MP populations, respectively.
For the MR population, the mean radial velocity and polar velocity ($\langle V_r\rangle$, $\langle V_\theta\rangle$) remain statistically consistent with zero throughout the radial range of 10 to 50~kpc. 
While the outer stellar halo (>~50~kpc) exhibits a net positive polar velocity, consistent with Figure~\ref{fig:single_component_result} and the prediction by \cite{2025MNRAS.544.2434S}.
Similar to Figure~\ref{fig:single_component_result}, the innermost MR stellar halo exhibits a strong prograde motion with $\langle V_\phi\rangle\sim-51~\mathrm{km~s^{-1}}$, which could be contamination from the thick disk. For the outer part of the MR stellar halo, the amplitude $\langle V_\phi\rangle$ is weaker than that in the single-component model, showing that the MR stellar halo does not exhibit a significant rotation at $9~\mathrm{kpc}<r_\text{GC}<30~\mathrm{kpc}$.

The velocity dispersion of the MR population also shows similar trends as in Figure~\ref{fig:single_component_result} above: the radial velocity dispersion $\sigma_r$ firstly reaches its peak at about 10~kpc and then roughly decreases with increasing $r_\mathrm{GC}$, reflecting the thick disk contamination for the innermost MR stellar halo. The azimuthal, polar velocity dispersion $\sigma_\phi$, $\sigma_\theta$ display an initial decline within $\sim$~30~kpc and are followed by a subsequent rise at larger radii. However, $\sigma_r$ is still a factor of twice larger than $\sigma_\phi$ and $\sigma_\theta$, even at $r_\mathrm{GC}>60~\mathrm{kpc}$, indicating a very radial stellar halo in the MR population even for the very outskirts.

The bottom panel of the left plot in Figure~\ref{fig:double_component_results} presents the measurements of the MP stellar halo. The mean velocities of the MP population show different behaviour compared with the MR population:\\
(1) The MP population represents a stable spin signal over a broader Galactocentric distance range of $3~\mathrm{kpc}<r_\mathrm{GC}<30~\mathrm{kpc}$. The mean azimuthal velocity reaches its most negative value of $\sim-30~\mathrm{km~s^{-1}}$ at $\sim$~20~kpc.
Given that the MP population is significantly more MP than the MW disk, we conclude that the net azimuthal motion observed in the MP stellar halo is very likely an intrinsic feature rather than contamination by disk stars. \\
(2) The MP population exhibits net radial and polar motions even in the inner stellar halo. Its mean polar velocity follows a distinct triphasic evolution: it first increases with galactocentric radius, reaches a peak of $\sim10~\mathrm{km~s^{-1}}$ at $\sim$~15~kpc, then decreases to near zero, and subsequently rises again at radii beyond $\sim$~30~kpc. The mean radial velocity of the MP population is roughly becoming more negative (inward motion) with the increase in radius, and the outer stellar halo exhibits stronger net inward motion ($r_\mathrm{GC}>50~\mathrm{kpc}$).

The net motion detected in the inner stellar halo may reflect the intrinsic motions of the progenitors associated with the multiple minor merger events.
The net motion in radial and polar velocities at $r_\mathrm{GC}>50~\mathrm{kpc}$ might reflect some perturbation from the LMC. We provide more discussions about the effect of LMC in Section~\ref{sec: lmc pertubation} below.

By comparing the mean azimuthal velocity of the MP and MR populations, we find that the MP population exhibits a higher azimuthal rotation than the MR population, and the difference exceeds 1$\sigma$ (see the third column in Table~\ref{table:double_component_results}). 
The typical $\langle V_\phi\rangle$ for the MP population\footnote{Here, for the MR population, we exclude the innermost bin ($3~\mathrm{kpc}<r_\mathrm{GC}<9~\mathrm{kpc}$), as it is contaminated by thick disk stars.} is about $-20~\mathrm{km~s^{-1}}$, and $-10~\mathrm{km~s^{-1}}$ for the MR population.
\cite{2017MNRAS.470.1259D} also reported that MR and MP populations exhibit different mean azimuthal velocity, but with an opposite trend than ours. The MR population ([Fe/H]>-1.5~dex) has a higher mean azimuthal velocity than the MP population ([Fe/H]<-1.5~dex) for various tracers (K giants, RR Lyrae, and BHBs). However, the difference in $\langle V_\phi\rangle$ between the two populations remains within the 1$\sigma$ uncertainty range in \cite{2017MNRAS.470.1259D}.

Using a local K-giant sample (heliocentric distance $<$ 4 kpc), \cite{2019ApJ...871..184T} found a non-monotonic trend with metallicity: the most MP halo stars (-2.5~dex<[Fe/H]<-1.6~dex) exhibit the strongest azimuthal rotation ($\langle V_\phi\rangle\sim-38~\mathrm{km~s^{-1}}$), whereas the intermediate-metallicity group (-1.6~dex<[Fe/H]<-1.3~dex) present the weakest azimuthal rotation ($\langle V_\phi\rangle\sim-11~\mathrm{km~s^{-1}}$), and the most MR halo (-1.3~dex<[Fe/H]<-1.0~dex) stars exhibit an amplitude between these motions ($\langle V_\phi\rangle\sim-24~\mathrm{km~s^{-1}}$). 
Although \citet{2021ApJ...919...66B} likewise detected a net azimuthal rotation of the stellar halo, they did not discuss the dependence on metallicity. And \cite{yr1rrlyrae} did not report an obvious net azimuthal rotation for both MP and MR stellar halo (see Table~6 in \cite{yr1rrlyrae}).

\begin{table*}[ht]
\centering
\caption{Best-fit parameters of the double-component model in different $r_\mathrm{GC}$ bins.}
\renewcommand{\arraystretch}{1.8}
\scriptsize      
\setlength{\tabcolsep}{3.4pt}  
\begin{tabular}{@{}lcccccccccccc@{}}
\toprule
$r_\mathrm{GC}$&$\langle V_r^\mathrm{MR}\rangle$ & $\langle V_\phi^\mathrm{MR}\rangle$ & $\langle V_\theta^\mathrm{MR}\rangle$ & $\sigma_r^\mathrm{MR}$ & $\sigma_\phi^\mathrm{MR}$ & $\sigma_\theta^\mathrm{MR}$ & $\mu_\mathrm{[Fe/H]}^\mathrm{MR}$& $\sigma_\mathrm{[Fe/H]}^\mathrm{MR}$ & $\beta_\mathrm{MR}$ & $p_\mathrm{MR}$\\
kpc& $\mathrm{km~s^{-1}}$ & $\mathrm{km~s^{-1}}$ & $\mathrm{km~s^{-1}}$ & $\mathrm{km~s^{-1}}$ & $\mathrm{km~s^{-1}}$ & $\mathrm{km~s^{-1}}$ & dex & dex &  &  \\
\midrule
$[3,160]$ & $0.9_{-0.8}^{+0.9}$ & $-7.1_{-0.3}^{+0.2}$ & $3.6_{-0.3}^{+0.4}$ & $131.4_{-0.6}^{+0.6}$ & $30.8_{-0.3}^{+0.3}$ & $35.0_{-0.2}^{+0.3}$ & $-1.146_{-0.001}^{+0.001}$ & $0.167_{-0.001}^{+0.001}$ & $0.937_{-0.002}^{+0.002}$ & $0.542_{-0.003}^{+0.003}$ & \\
$[3,9)$ & $0.3_{-2.8}^{+3.2}$ & $-51.1_{-1.8}^{+1.9}$ & $-0.9_{-1.0}^{+1.4}$ & $153.5_{-2.1}^{+2.2}$ & $88.4_{-1.4}^{+1.3}$ & $60.5_{-1.0}^{+0.9}$ & $-1.075_{-0.005}^{+0.005}$ & $0.146_{-0.003}^{+0.004}$ & $0.756_{-0.013}^{+0.013}$ & $0.517_{-0.013}^{+0.015}$ &\\
$[9,11.5)$ & $4.4_{-2.3}^{+3.0}$ & $-9.6_{-1.1}^{+0.9}$ & $1.9_{-0.9}^{+0.8}$ & $164.9_{-2.5}^{+2.5}$ & $40.3_{-1.4}^{+1.3}$ & $43.4_{-0.8}^{+0.8}$ & $-1.148_{-0.004}^{+0.004}$ & $0.182_{-0.003}^{+0.005}$ & $0.935_{-0.004}^{+0.004}$ & $0.524_{-0.012}^{+0.012}$ &\\
$[11.5,13.7)$ & $0.6_{-3.1}^{+2.7}$ & $-6.6_{-0.8}^{+0.8}$ & $4.0_{-0.9}^{+0.9}$ & $146.9_{-2.5}^{+2.1}$ & $30.0_{-0.8}^{+0.8}$ & $38.1_{-0.7}^{+0.7}$ & $-1.152_{-0.004}^{+0.004}$ & $0.175_{-0.004}^{+0.004}$ & $0.946_{-0.004}^{+0.004}$ & $0.568_{-0.008}^{+0.009}$ &\\
$[13.7,15.9)$ & $-2.9_{-2.6}^{+2.5}$ & $-8.4_{-0.7}^{+0.9}$ & $1.3_{-0.7}^{+0.9}$ & $140.2_{-1.7}^{+1.8}$ & $29.0_{-1.0}^{+0.9}$ & $33.6_{-0.8}^{+0.9}$ & $-1.154_{-0.005}^{+0.005}$ & $0.174_{-0.004}^{+0.003}$ & $0.950_{-0.004}^{+0.004}$ & $0.581_{-0.010}^{+0.010}$ &\\
$[15.9,18.2)$ & $3.8_{-2.6}^{+2.8}$ & $-8.2_{-0.7}^{+0.7}$ & $3.7_{-0.7}^{+0.8}$ & $132.8_{-1.3}^{+1.7}$ & $25.5_{-0.8}^{+0.8}$ & $31.2_{-0.6}^{+0.8}$ & $-1.146_{-0.005}^{+0.004}$ & $0.169_{-0.003}^{+0.003}$ & $0.954_{-0.003}^{+0.004}$ & $0.592_{-0.010}^{+0.011}$ &\\
$[18.2,20.8)$ & $0.3_{-3.3}^{+1.8}$ & $-7.7_{-0.9}^{+0.8}$ & $6.4_{-1.1}^{+0.7}$ & $121.8_{-2.2}^{+1.3}$ & $25.3_{-0.8}^{+1.0}$ & $30.0_{-0.7}^{+0.9}$ & $-1.161_{-0.004}^{+0.005}$ & $0.177_{-0.003}^{+0.005}$ & $0.948_{-0.004}^{+0.004}$ & $0.663_{-0.015}^{+0.010}$ &\\
$[20.8,24.2)$ & $0.9_{-1.9}^{+2.1}$ & $-6.3_{-0.8}^{+0.7}$ & $4.8_{-0.7}^{+0.7}$ & $114.5_{-1.3}^{+1.2}$ & $24.0_{-0.8}^{+0.8}$ & $28.4_{-0.7}^{+0.7}$ & $-1.156_{-0.004}^{+0.004}$ & $0.173_{-0.003}^{+0.004}$ & $0.947_{-0.004}^{+0.004}$ & $0.644_{-0.010}^{+0.010}$ &\\
$[24.2,28.7)$ & $3.8_{-2.0}^{+1.9}$ & $-4.1_{-0.8}^{+0.9}$ & $4.6_{-1.0}^{+0.8}$ & $107.3_{-1.5}^{+1.7}$ & $25.7_{-0.8}^{+0.9}$ & $25.1_{-0.9}^{+0.8}$ & $-1.152_{-0.004}^{+0.005}$ & $0.165_{-0.004}^{+0.003}$ & $0.944_{-0.005}^{+0.005}$ & $0.618_{-0.010}^{+0.010}$ &\\
$[28.7,35.8)$ & $1.4_{-2.4}^{+2.4}$ & $-1.9_{-1.1}^{+1.0}$ & $4.1_{-1.0}^{+1.1}$ & $106.8_{-1.5}^{+1.5}$ & $28.0_{-1.4}^{+1.6}$ & $29.6_{-1.3}^{+1.3}$ & $-1.146_{-0.005}^{+0.006}$ & $0.150_{-0.004}^{+0.004}$ & $0.927_{-0.009}^{+0.009}$ & $0.565_{-0.011}^{+0.012}$ &\\
$[35.8,47.5)$ & $-0.7_{-2.5}^{+2.2}$ & $-4.2_{1.4}^{+1.5}$ & $3.8_{-1.4}^{+1.2}$ & $105.8_{-1.9}^{+1.6}$ & $33.6_{-2.0}^{+1.9}$ & $35.1_{-2.0}^{+1.8}$ & $-1.145_{-0.007}^{+0.013}$ & $0.143_{-0.005}^{+0.005}$ & $0.895_{-0.016}^{+0.014}$ & $0.499_{-0.013}^{+0.014}$ &\\
$[47.5,60)$ & $-0.5_{-4.9}^{+5.2}$ & $-5.4_{-4.3}^{+3.1}$ & $0.5_{-4.3}^{+3.9}$ & $101.9_{-3.0}^{+3.7}$ & $37.2_{-4.4}^{+4.8}$ & $44.1_{-4.4}^{+5.3}$ & $-1.117_{-0.012}^{+0.022}$ & $0.123_{-0.011}^{+0.013}$ & $0.840_{-0.044}^{+0.054}$ & $0.401_{-0.031}^{+0.031}$ &\\
$[60,80)$ & $-0.4_{-7.9}^{+8.8}$ & $-3.3_{-4.7}^{+5.1}$ & $15.1_{-5.6}^{+4.8}$ & $105.6_{-4.6}^{+6.1}$ & $28.4_{-10.1}^{+8.1}$ & $46.8_{-6.2}^{+6.3}$ & $-1.141_{-0.019}^{+0.094}$ & $0.135_{-0.016}^{+0.019}$ & $0.866_{-0.054}^{+0.058}$ & $0.435_{-0.045}^{+0.044}$ &\\
$[80,160]$ & $9.8_{-17.1}^{+19.0}$ & $14.1_{-14.0}^{+13.3}$ & $17.9_{-21.6}^{+12.6}$ & $99.8_{-8.6}^{+16.2}$ & $51.1_{-21.4}^{+80.7}$ & $45.3_{-29.6}^{+141.2}$ & $-1.345_{-0.062}^{+0.071}$ & $0.280_{-0.100}^{+0.096}$ & $0.766_{-0.33}^{+1.44}$ & $0.653_{-0.436}^{+0.180}$ &\\
\midrule
$r_\mathrm{GC}$&$\langle V_r^\mathrm{MP}\rangle$ & $\langle V_\phi^\mathrm{MP}\rangle$ & $\langle V_\theta^\mathrm{MP}\rangle$ & $\sigma_r^\mathrm{MP}$ & $\sigma_\phi^\mathrm{MP}$ & $\sigma_\theta^\mathrm{MP}$ & $\mu_\mathrm{[Fe/H]}^\mathrm{MP}$& $\sigma_\mathrm{[Fe/H]}^\mathrm{MP}$ & $\beta_\mathrm{MP}$\\
kpc& $\mathrm{km~s^{-1}}$ & $\mathrm{km~s^{-1}}$ & $\mathrm{km~s^{-1}}$ & $\mathrm{km~s^{-1}}$ & $\mathrm{km~s^{-1}}$ & $\mathrm{km~s^{-1}}$ & dex & dex & \\
\midrule
$[3,160]$ & $-2.3_{-0.8}^{+1.0}$ & $-22.8_{-0.7}^{+0.6}$ & $4.4_{-0.6}^{+0.7}$ & $131.3_{-0.7}^{+0.7}$ & $103.2_{-0.6}^{+0.6}$ & $92.2_{-0.4}^{+0.5}$ & $-1.578_{-0.003}^{+0.003}$ & $0.429_{-0.002}^{+0.002}$ & $0.445_{-0.012}^{+0.012}$ &\\
$[3,9)$ & $7.6_{-2.8}^{+3.1}$ & $-16.6_{-2.0}^{+2.1}$ & $2.8_{-1.8}^{+1.8}$ & $145.5_{-2.3}^{+2.1}$ & $104.2_{-1.5}^{+1.6}$ & $94.3_{-1.2}^{+1.5}$ & $-1.615_{-0.013}^{+0.011}$ & $0.387_{-0.005}^{+0.006}$ & $0.533_{-0.026}^{+0.028}$ &\\
$[9,11.5)$ & $1.9_{-3.2}^{+2.6}$ & $-20.9_{-2.4}^{+1.9}$ & $5.0_{-1.9}^{+1.8}$ & $145.1_{-1.9}^{+1.9}$ & $106.0_{-1.4}^{+1.9}$ & $93.4_{-1.4}^{+1.5}$ & $-1.617_{-0.014}^{+0.011}$ & $0.439_{-0.007}^{+0.007}$ & $0.526_{-0.025}^{+0.027}$ &\\
$[11.5,13.7)$ & $-3.5_{-2.9}^{+3.3}$ & $-16.4_{-2.2}^{+2.6}$ & $8.0_{-1.9}^{+1.7}$ & $142.1_{-2.4}^{+2.1}$ & $100.4_{-1.7}^{+1.7}$ & $90.3_{-1.5}^{+1.7}$ & $-1.646_{-0.011}^{+0.011}$ & $0.436_{-0.007}^{+0.007}$ & $0.548_{-0.030}^{+0.029}$ &\\
$[13.7,15.9)$ & $-2,9_{-2.6}^{+3.1}$ & $-23.0_{-1.8}^{+2.2}$ & $9.9_{-2.1}^{+2.0}$ & $135.4_{-2.3}^{+2.3}$ & $96.5_{-1.4}^{+1.5}$ & $89.6_{-1.5}^{+1.5}$ & $-1.605_{-0.009}^{+0.009}$ & $0.407_{-0.005}^{+0.006}$ & $0.528_{-0.032}^{+0.031}$ &\\
$[15.9,18.2)$ & $-3.1_{-2.8}^{+3.1}$ & $-26.3_{-2.0}^{+2.2}$ & $5.7_{-1.9}^{+1.7}$ & $126.9_{-1.9}^{+1.9}$ & $98.7_{-1.7}^{+2.0}$ & $82.1_{-1.4}^{+1.2}$ & $-1.586_{-0.009}^{+0.008}$ & $0.415_{-0.006}^{+0.007}$ & $0.488_{-0.031}^{+0.029}$ &\\
$[18.2,20.8)$ & $-4.6_{-3.9}^{+5.6}$ & $-28.8_{-2.6}^{+3.6}$ & $2.9_{-3.0}^{+2.2}$ & $127.1_{-2.0}^{+2.1}$ & $99.1_{-2.8}^{+2.0}$ & $90.8_{-2.6}^{+1.8}$ & $-1.599_{-0.015}^{+0.017}$ & $0.451_{-0.013}^{+0.007}$ & $0.440_{-0.047}^{+0.040}$ &\\
$[20.8,24.2)$ & $-4.6_{-3.1}^{+3.0}$ & $-20.2_{-2.8}^{+2.3}$ & $3.4_{-2.2}^{+2.3}$ & $129.5_{-2.4}^{+2.3}$ & $99.8_{-1.9}^{+2.1}$ & $93.3_{-2.0}^{2.0}$ & $-1.598_{-0.012}^{+0.012}$ & $0.438_{-0.007}^{+0.009}$ & $0.443_{-0.043}^{+0.042}$ &\\
$[24.2,28.7)$ & $-7.7_{-2.9}^{+2.7}$ & $-17.9_{-2.9}^{+2.8}$ & $0.7_{-2.6}^{+2.4}$ & $124.8_{-1.9}^{+2.3}$ & $98.5_{-1.8}^{+2.2}$ & $92.6_{-1.7}^{+2.0}$ & $-1.592_{-0.011}^{+0.012}$ & $0.412_{-0.007}^{+0.007}$ & $0.414_{-0.038}^{+0.046}$ &\\
$[28.7,35.8)$ & $-12.4_{-3.6}^{+3.5}$ & $-11.1_{-2.8}^{+2.5}$ & $2.9_{-2.9}^{+2.8}$ & $121.8_{-2.1}^{+2.4}$ & $100.9_{-2.3}^{+2.3}$ & $104.7_{-2.2}^{+2.4}$ & $-1.636_{-0.012}^{+0.012}$ & $0.418_{-0.007}^{+0.006}$ & $0.288_{-0.055}^{+0.062}$ &\\
$[35.8,47.5)$ & $-2.6_{-3.5}^{+2.5}$ & $-8.6_{-3.0}^{+2.5}$ & $4.4_{-2.9}^{+2.6}$ & $115.5_{-1.9}^{+2.1}$ & $107.7_{-2.6}^{+2.5}$ & $112.1_{-2.6}^{+2.3}$ & $-1.628_{-0.014}^{+0.012}$ & $0.408_{-0.008}^{+0.007}$ & $0.094_{-0.073}^{+0.070}$ &\\
$[47.5,60)$ & $-9.1_{-3.7}^{+4.4}$ & $6.5_{-3.6}^{+3.9}$ & $7.1_{-5.2}^{+4.4}$ & $109.8_{-3.1}^{+3.0}$ & $106.9_{-3.7}^{+3.9}$ & $115.3_{-3.6}^{+4.4}$ & $-1.625_{-0.023}^{+0.024}$ & $0.395_{-0.012}^{+0.011}$ & $-0.025_{-0.123}^{+0.138}$ &\\
$[60,80)$ & $-14.5_{-5.9}^{+6.2}$ & $2.4_{-6.1}^{+7.2}$ & $15.3_{-7.4}^{+8.3}$ & $105.3_{-3.9}^{+4.0}$ & $102.0_{-5.1}^{+5.9}$ & $125.6_{-6.9}^{+7.3}$ & $-1.625_{-0.031}^{+0.036}$ & $0.384_{-0.016}^{+0.017}$ & $-0.180_{-0.229}^{+0.238}$ &\\
$[80,160]$ & $0.6_{-22.3}^{+27.0}$ & $17.0_{-36.2}^{+22.1}$ & $12.8_{-36.5}^{+16.7}$ & $108.2_{-15.8}^{+13.1}$ & $130.5_{-54.6}^{+35.9}$ & $141.1_{-65.9}^{+69.9}$ & $-1.600_{-0.152}^{+0.141}$ & $0.443_{-0.079}^{+0.102}$ & $-0.577_{-1.99}^{+2.00}$ &\\
\bottomrule
\end{tabular}
\label{table:double_component_results}
\end{table*}

Furthermore, in Figure~\ref{fig:double_component_results}, the azimuthal and polar velocity dispersion profiles of the MP population are markedly discrepant from those of the MR population, reinforcing a pronounced chemo-dynamical segregation. $\sigma_r$ of the MP population is a monotonic function of distance, without exhibiting a rapid increase in the inner region as that in the MR population. $\sigma_\phi$ is nearly a constant ($\sim100~\mathrm{km~s^{-1}}$) from 3 to 80~kpc, and increases only for the outermost bin, and $\sigma_\phi$ only increases beyond 40~kpc. $\sigma_\phi$ and $\sigma_\phi$ also exhibit greater amplitudes than those in the MR population, indicating that the MP population is more isotropic than the MR population.

\subsubsection{Radial dependent velocity anisotropy of metal-rich and metal-poor components}

The right plot of Figure~\ref{fig:double_component_results} presents the velocity anisotropy dependence on Galactocentric distance of the MR (top) and MP (bottom) populations. In comparison, we also plot the velocity anisotropy profiles based on the single-component model and measurements from \cite{2021ApJ...919...66B} and \cite{yr1rrlyrae} (the black, orange, and green dots, see Section~\ref{sec:single_component_results} and Figure~\ref{fig:single_component_result}). 
The MR population is highly radial with $0.75<\beta_\mathrm{MR}<0.96$ over $3~\mathrm{kpc}<r_\mathrm{GC}<100~\mathrm{kpc}$. $\beta_\mathrm{MR}$ firstly rises from $\sim$~0.75 to $\sim$~0.92 within the inner 10~kpc, and remains approximately constant ($\sim0.90$) out to 50~kpc, which then exhibits a modest decline beyond 50~kpc. $\beta_\mathrm{MP}$ roughly declines monotonically with $r_\mathrm{GC}$. The rising trend with increasing $r_\mathrm{GC}$ and within 10~kpc might be due to the contamination of some thick disk stars at smaller radii, hence causing $\beta$ slightly more tangential. The MP population approaches isotropy near 50~kpc and becomes tangentially biased at larger radii. Although the velocity anisotropy profile in the outer stellar halo could have been overestimated due to distance uncertainties, it only affects the profile beyond $\sim$~70~kpc, where the $\beta$ measurements already have large errorbars (see Appendix~\ref{app:error_on_model} for details).

The combination of the radially varying anisotropy profiles of MR and MP populations can naturally reproduce the anisotropy profile (the black dots) fit with the full sample, and simultaneously reconciles the apparent discrepancies among the three literature studies displayed in Figure~\ref{fig:single_component_result}. 
For example, although the velocity anisotropy of the MR population remains approximately constant at $20~\mathrm{kpc}<r_\mathrm{GC}<50~\mathrm{kpc}$, due to the decreasing fraction of the MR population ($P_\mathrm{MR}$) and the decreasing anisotropy profile of the MP population, the final mixed velocity anisotropy is decreasing at this range. 

Although different survey observed samples cover similar [Fe/H] distribution ($\sim$~-3.5 to $\sim$~-1 dex, see Figure~1 in \cite{2021ApJ...919...66B}, Figure~14 in \cite{yr1rrlyrae}, the different velocity anisotropy profiles observed in this work, \cite{2021ApJ...919...66B}, and \cite{yr1rrlyrae} could be explained by the different MR population fractions. 
In \cite{2021ApJ...919...66B}, the fraction of the MR population  (-1.7~dex<[Fe/H]<-1.0~dex) for their K giant sample is about 62~\% (see Figure~5 in \cite{2021ApJ...919...66B}), greater than in this paper, hence may lead to a higher velocity anisotropy beyond $\sim$~30~kpc. 
The fraction of the MR population for the BHB sample in \cite{2021ApJ...919...66B} is about 28~\%, hence exhibits a significantly lower velocity anisotropy than their K giant sample. And the velocity anisotropy profile is more consistent with our MP population results (cyan dots connected by solid lines in the bottom panel of Figure~\ref{fig:double_component_results}).
In \cite{yr1rrlyrae}, the fraction of the MR population is smaller than that in this paper (see Table~6 and Figure~16 in \cite{yr1rrlyrae}), which may explain why \cite{yr1rrlyrae} presents a lower velocity anisotropy than our single-component model prediction at 10~kpc to 30~kpc.

The MR and MP populations exhibit markedly distinct chemo-kinematic signatures, characterized by divergent velocity anisotropies, dispersion profiles, and mean velocity profiles that collectively might reflect that they have very different formation and accretion histories. The MR population is likely mainly contributed by the GSE merger event, as the progenitor of GSE fell in with a very radial orbit \citep{2018MNRAS.478..611B,2018Natur.563...85H}. Figures~\ref{fig:mcmc_2components} and ~\ref{fig:fraction_of_metal_rich} might also suggest that the GSE remnant could even extend to $\sim$~70-80~kpc. Because the MR population still contributes $\sim40\%$ to the total stellar halo for the very outer stellar halo, and with very radial orbits. While the MP population might be associated with multiple minor merger events, according to its broader metallicity distribution. The radially dependent velocity anisotropy profile of the MP component may indicate that the progenitors of these multiple minor merger events accreted onto the MW with distinct orbits and velocity anisotropies.

\subsection{LMC Perturbed Velocity Field}
\label{sec: lmc pertubation}

In Sections~\ref{sec:single_component_results}, ~\ref{sec:doubel_components_results1} and \ref{sec:doubel_components_results2}, we find that the outer stellar halo ($r_\mathrm{GC}>50~\mathrm{kpc}$) exhibits non-zero mean velocity in radial and polar velocities ($V_r$ and $V_\theta$). According to MW-LMC simulations \citep{2019ApJ...884...51G,2024MNRAS.534.2694S,2024MNRAS.527..437V,2024MNRAS.531.3524Y,2025MNRAS.544.2434S,2025MNRAS.544.1820Y} or rigid MW-LMC simulations with stellar halo particles \citep{2026MNRAS.545f2111B}, the infall of the LMC imprints a pronounced perturbation on the stellar halo radial velocity distribution due to the reflex motion of the MW disk and inner halo, resulting in a global positive bias in polar velocity distribution. These perturbations induced by LMC are observable at $r_\mathrm{GC}>30~\mathrm{kpc}$ and become more prominent beyond $\sim$~50-60~kpc.  
The perturbed radial velocity distribution appears to be a net positive mean radial velocity in the northern stellar halo, and a nearly equal amplitude negative mean radial velocity in the southern halo (see Figure~4 in \cite{2024MNRAS.527..437V} and Figure~4 in \cite{2025MNRAS.544.2434S}), which forms a dipole field. 

In this section, we quantify the reflex motion of the MW, 
we first select the halo K giants with a constraint solely on the radial velocity deviation, restricting observational uncertainty of radial velocity to be $|\delta V_r|<50~\mathrm{km~s^{-1}}$, while removing any constraints on the azimuthal $|\delta V_\phi|$, and polar $|\delta V_\theta|$ velocity uncertainties \footnote{As we only focus on the radial velocity distribution, and we have already found the global positive polar velocity induced by LMC in Section~\ref{sec:single_component_results} and Section~\ref{sec:doubel_components_results2}, and the infalling of LMC does not change the azimuthal velocity distribution of the stellar halo.}. 

\begin{figure}
\begin{center}
\includegraphics[width=0.49\textwidth]{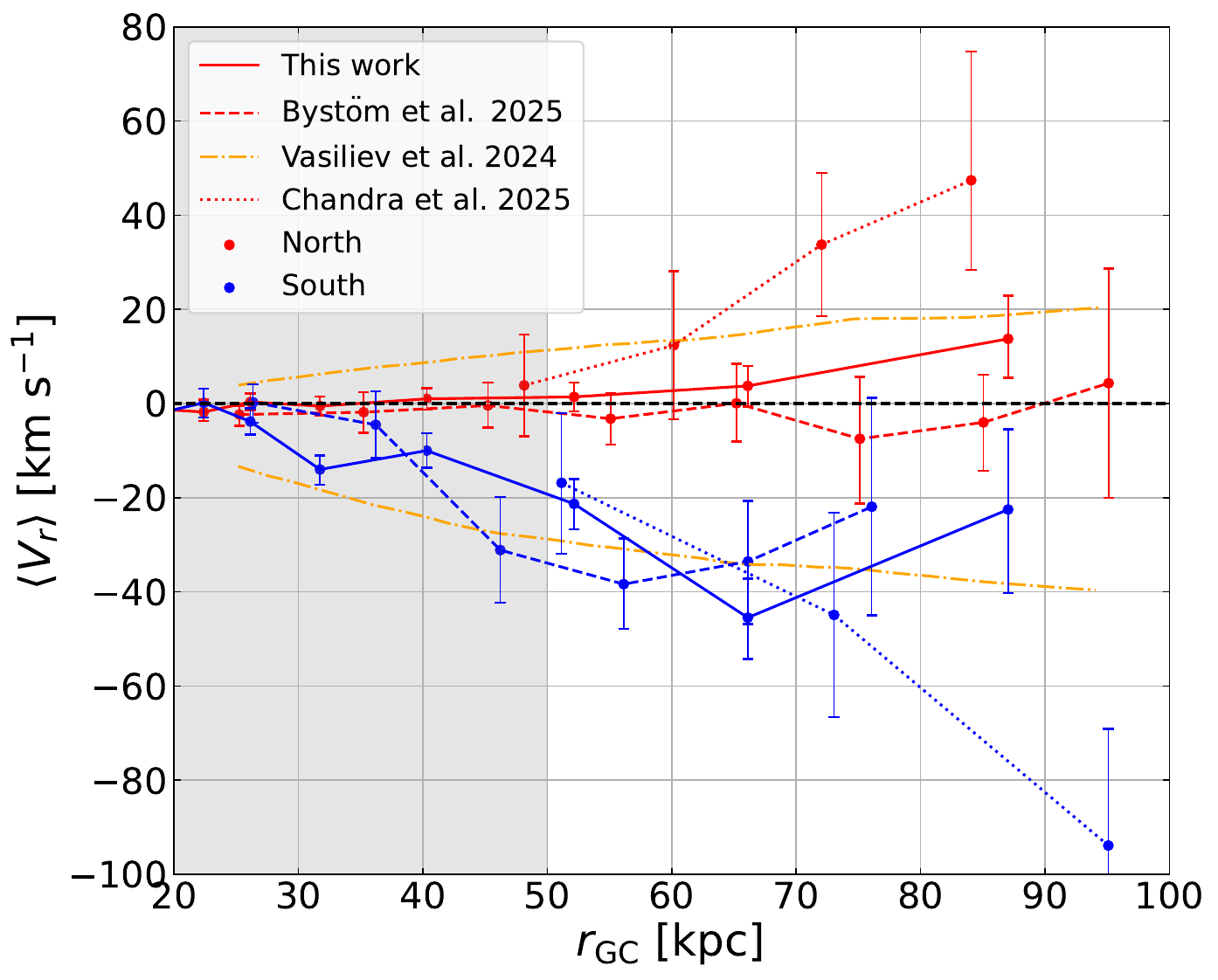}%
\end{center}
\caption{The reflex motion of the outer stellar halo quantified by the mean radial-velocity profiles of the northern (red) and southern (blue) stellar haloes. The solid, dashed, and dotted lines present the measurements in this work (based on K giants), those in \cite{2025MNRAS.542..560B} (based on BHB stars), and in \cite{2025ApJ...988..156C} (based on red giants), respectively. The orange dash-dotted line shows the predictions from \cite{2024MNRAS.527..437V} based on the L2-M11 model (see \cite{2024MNRAS.527..437V} for details).
The black dashed line indicates where $\langle V_r\rangle=0~\mathrm{km~s^{-1}}$.}
\label{fig:mean_vr_profile}
\end{figure}

\begin{figure}
\begin{center}
\includegraphics[width=0.49\textwidth]{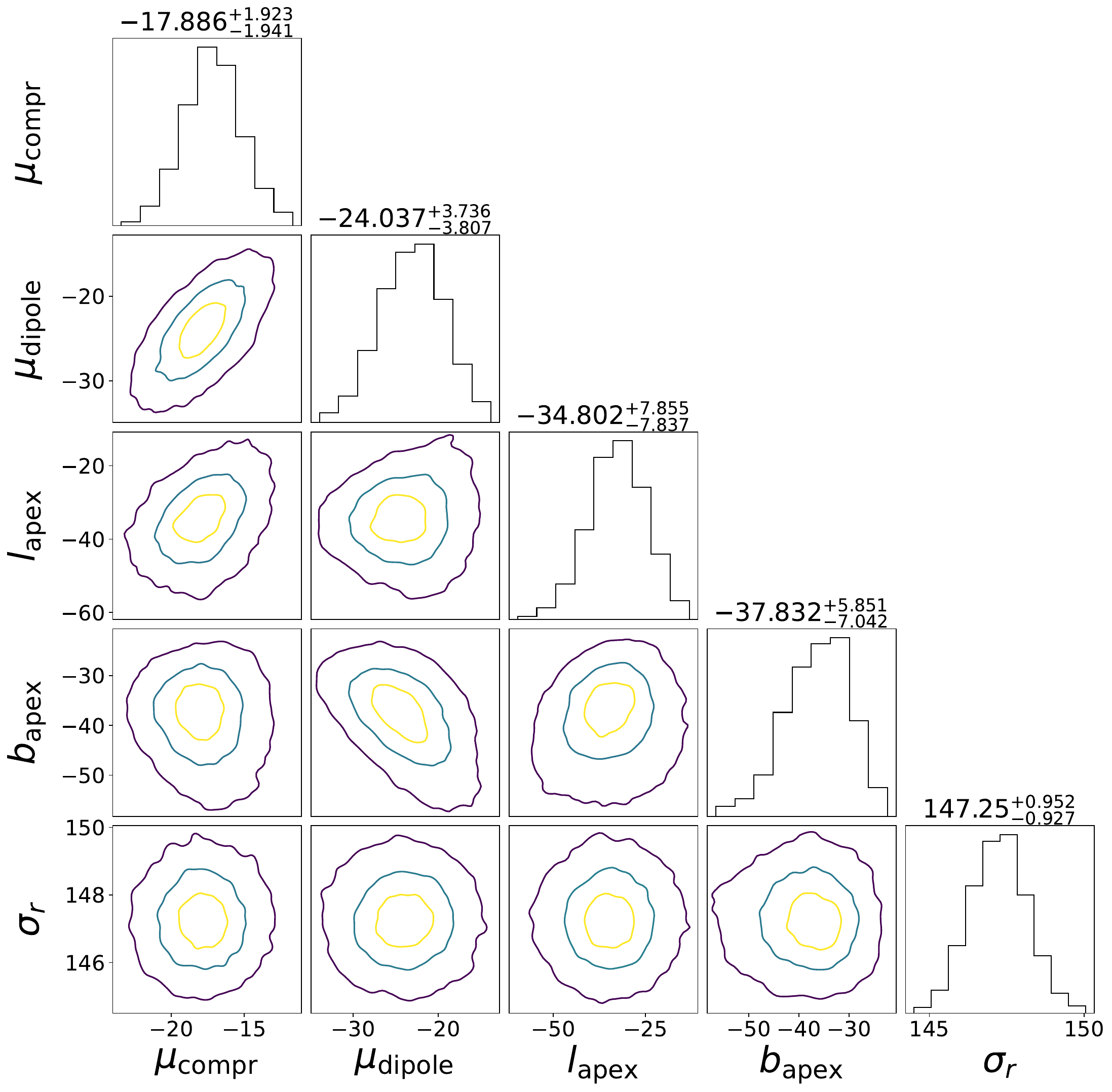}%
\end{center}
\caption{Error contours for different parameter combinations of the dipole velocity field model (see Equation~\ref{equ:LMC_mean_velocity_field}). $\mu_\mathrm{compr}$ represents the overall motion of the outer stellar halo, and $\mu_\mathrm{dipole}$ represents the amplitude of a dipole velocity field. $l_\mathrm{apex}, b_\mathrm{apex}$ stand for the dipole apex direction. $\sigma_r$ shows the velocity dispersion in the radial velocity distribution.}
\label{fig:mcmc_lmc}
\end{figure}

The reflex motion of the stellar halo can be quantified by:\\ 
(1) Measuring the mean velocity of the northern and southern stellar halo at different distance bins.\\
(2) Fitting the mean velocity at different sky positions with a dipole velocity field model (see Section~\ref{sec:dipole_velocity_field_model}, Equation~\ref{equ:LMC_perturbation_likelihood} and Equation~\ref{equ:LMC_mean_velocity_field}).

\begin{figure*}
\begin{center}
\includegraphics[width=0.99\textwidth]{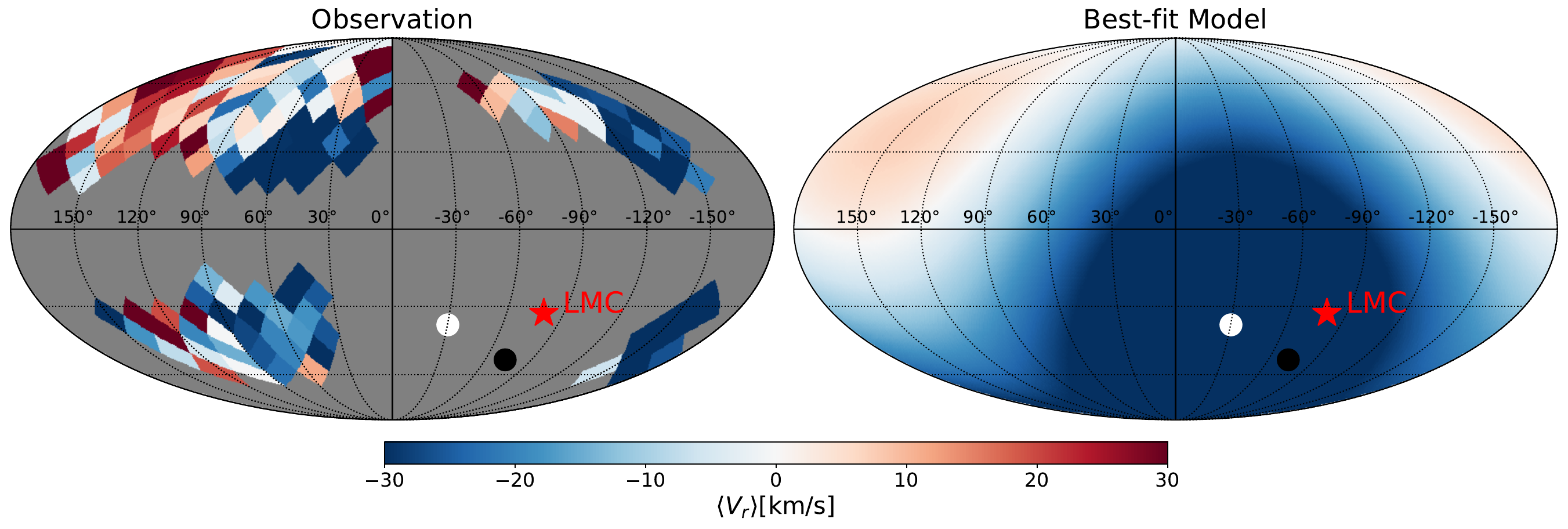}%
\end{center}
\caption{The mean radial velocity from observation (left panel) and best-fit model (right panel) in Mollweide projection. The observation covers a distance range of $50~\mathrm{kpc}<r_\mathrm{GC}<120~\mathrm{kpc}$. The best-fit model is computed by fitting the observation with Equation~\ref{equ:LMC_mean_velocity_field}. The apex direction of the best-fit dipole velocity field derived in this work is shown in white, while that of \cite{2025MNRAS.542..560B}}. The red pentagram shows the current position of LMC.
\label{fig:LMC_overall_perturbation}
\end{figure*}

We first fit the mean radial velocity $\langle V_r\rangle$ profile as a function of distance for the northern (Galactic latitude $b > 0$) and southern stellar halo ($b < 0$). Figure~\ref{fig:mean_vr_profile} presents the mean radial velocity of the northern (blue lines) and southern (red lines) stellar halo as a function of Galactocentric distance. 
The mean velocity is fit by assuming that the radial velocity is a Gaussian distribution\footnote{In this section, we also assume that the MR and MP populations share a similar radial velocity distribution, hence the overall distribution can be described by a single Gaussian distribution.} ($x=V_r$, see Equation~\ref{equ:single_guassian}). The dashed lines present the measurements in \cite{2025MNRAS.542..560B} based on a BHB sample \footnote{\cite{2025MNRAS.542..560B} measured the mean Galactic Standard of Rest radial velocity ($V_\text{GSR}$) across different Galactocentric radii, which is Heliocentric. For sufficiently large $r_\text{GC}$, $V_r\approx V_\text{GSR}$. Hence  $\langle V_r\rangle \approx \langle V_\text{GSR}\rangle$.}, the dotted lines show the results measured in \cite{2025ApJ...988..156C} with a red giant sample\footnote{\cite{2025ApJ...988..156C} compiled spectroscopic observations from the H3, SEGUE, and MagE surveys to construct their all-sky sample of luminous red giant stars in the Milky Way’s outer halo. As illustrated in their Figure 2, the MagE survey provides the dominant contribution of red giant stars for Galactocentric radii greater than 50 kpc},
and the solid lines stand for the results in this work with our K giant sample. Due to the larger sample size of K giants than BHBs, the errorbars in our measurements are smaller than those of \cite{2025MNRAS.542..560B}. In comparison, we also plot the prediction from the N-body simulation; the mean radial velocity for the northern and southern stellar halo is from the L2-M11 model\footnote{Similar to \cite{2025MNRAS.542..560B}, to make a robust comparisons between the model and the data, we apply the DESI footprint cut to the model.} in \cite{2024MNRAS.527..437V}.
The shaded region denotes $r_\mathrm{GC}<50~\mathrm{kpc}$, where the perturbation from LMC is expected to be less pronounced.

The mean radial velocity profiles measured in this work slightly differ from those in \cite{2025MNRAS.542..560B}. First, we find that for the northern stellar halo, the mean radial velocity in the inner region ($r_\mathrm{GC}<50~\mathrm{kpc}$) is consistent with zero. For the outer part ($r_\mathrm{GC}>50~\mathrm{kpc}$), the mean radial velocity increases with distance in our measurement, whereas the measurements of the northern stellar halo in \cite{2025MNRAS.542..560B} do not exhibit a monotonic increase. On the other hand, for the southern stellar halo, both measurements show a non-monotonic trend. The mean radial velocity in the southern stellar halo first increases with distance, reaching the most negative peak, and subsequently decreases again, though the errorbars are quite large beyond 70~kpc. However, the mean radial velocity reaches its most negative peak at $\sim$~66~kpc in this work and $\sim$~56~kpc in \cite{2025MNRAS.542..560B}, but the difference between the two measurements is still within 1$\sigma$ uncertainties. Hence, we conclude that our measurements are quite consistent with \cite{2025MNRAS.542..560B}.

The mean radial velocity profiles measured in \cite{2025ApJ...988..156C} exhibit a more pronounced discrepancy compared with this work and \cite{2025MNRAS.542..560B}. The northern and southern radial velocity profiles presented in \cite{2025ApJ...988..156C} both rise monotonically with radius. The northern profile reveals a significantly stronger bulk motion signal than that derived in this work, whereas the southern profile is in far better agreement with our results\footnote{We also note that \cite{2025ApJ...988..156C} only adopts the LMC Quadrant ($l<0$ and $b<0$) and the opposite Quadrant ($l>0$ and $b>0$) for computing the mean radial velocity profile (see Figure~3 and Figure~4 in this paper), differs from our choice, which will also contribute to the observed discrepancy.}.
This discrepancy may arise from the observational footprint differences between our sample and that in \cite{2025ApJ...988..156C}, which can be seen from the left panel of Figure~\ref{fig:LMC_overall_perturbation} and Figure~2 in \cite{2025ApJ...988..156C}. As shown in \cite{2025arXiv251004735B}, the DESI footprint `flattens' the $\langle V_r\rangle$ signal in the north as it is averaged over regions of net positive/negative signal.

Figure~\ref{fig:mean_vr_profile} also reveals that the southern halo attains a markedly larger mean radial velocity amplitude than its northern counterpart. This contrasts with the prediction of \cite{2024MNRAS.527..437V}, whose model yields an obvious mean radial velocity in the northern halo (orange dash–dot line in the same figure). Similar to our results, \cite{2025MNRAS.542..560B} also reported a significantly stronger radial-velocity signal in the southern halo, while the northern halo does not exhibit obvious net motion.

Following \cite{2025MNRAS.542..560B}, we then fit the mean radial velocity of the outer stellar halo ($50~\mathrm{kpc}<r_\mathrm{GC}<120~\mathrm{kpc}$) as a function of sky position with the dipole velocity field model (see Section~\ref{sec:dipole_velocity_field_model}). The contracting motion of the outer stellar is also fit in this model. We first fit the dipole radial velocity field using all K giants with $50~\mathrm{kpc}<r_\mathrm{GC}<120~\mathrm{kpc}$, and then present the best-fit results at a few distance bins (see the first column in Table~\ref{table:LMC_results}).

Figure~\ref{fig:mcmc_lmc} presents the error contours for different parameter combinations and the marginalized posterior distribution for each parameter in Equation~\ref{equ:LMC_mean_velocity_field} using halo stars at a distance range of $50~\mathrm{kpc}<r_\mathrm{GC}<120~\mathrm{kpc}$. The best-fit parameters and 1$\sigma$ uncertainties are presented in the first row of Table~\ref{table:LMC_results}. We find that: (1) The outer stellar halo is undergoing a contracting motion with $\mu_\mathrm{compr}\sim-18~\mathrm{km~s^{-1}}$. The net inward contraction of the outer halo superposed on the reflex motion induced by the descending disk, systematically boosting the apparent infall velocity in the south while weakening it in the north, thereby accounting for the larger mean radial-velocity amplitude observed in the southern stellar halo and the smaller mean radial-velocity amplitude observed in the northern stellar halo (see Figure~\ref{fig:mean_vr_profile}). (2) The amplitude of the dipole velocity field of the reflex motion ($\mu_\mathrm{dipole}$) is about -24$~\mathrm{km~s^{-1}}$, showing a slightly larger amplitude with the contracting motion, and is consistent with the predictions of the N-body simulation.

Similarly, \cite{2025MNRAS.542..560B} and \cite{2025ApJ...988..156C} reported that the outer stellar halo is moving inwards with $\mu_\mathrm{compr}$ of $\sim-24~\mathrm{km~s^{-1}}$ and $\sim-12~\mathrm{km~s^{-1}}$\footnote{See the middle panel of Figure~3 in \cite{2025ApJ...988..156C}.}, respectively. \cite{2025MNRAS.542..560B} also reported that the amplitude of the dipole velocity field is slightly greater than the dipole velocity of the reflex motion  ($\mu_\mathrm{dipole}\sim-34~\mathrm{km~s^{-1}}$ in their measurement). The difference between $\mu_\mathrm{compr}$ and $\mu_\mathrm{dipole}$ in this work and those in \cite{2025MNRAS.542..560B} is within 1$\sigma$ uncertainty though. However, the position of the apex of the dipole velocity field is not fully consistent. We report $l_\mathrm{apex}=-30°,b_\mathrm{apex}=-38°$, while \cite{2025MNRAS.542..560B} find that $l_\mathrm{apex}=-73°,b_\mathrm{apex}=-53°$ (see Table~1 in \cite{2025MNRAS.542..560B}).
The two independent measurements of the $l_\mathrm{apex}$ differ by approximately 2.4 $\sigma$, whereas the two measurements of the $b_\mathrm{apex}$ differ by $\sim$~1.1 $\sigma$. The discrepancy might be explained by the fact that our sample contains more distant stars than that in \cite{2025MNRAS.542..560B}. As we will discuss below, the parameters of the dipole velocity field exhibit some dependence on radii.

\begin{figure}
\begin{center}
\includegraphics[width=0.49\textwidth]{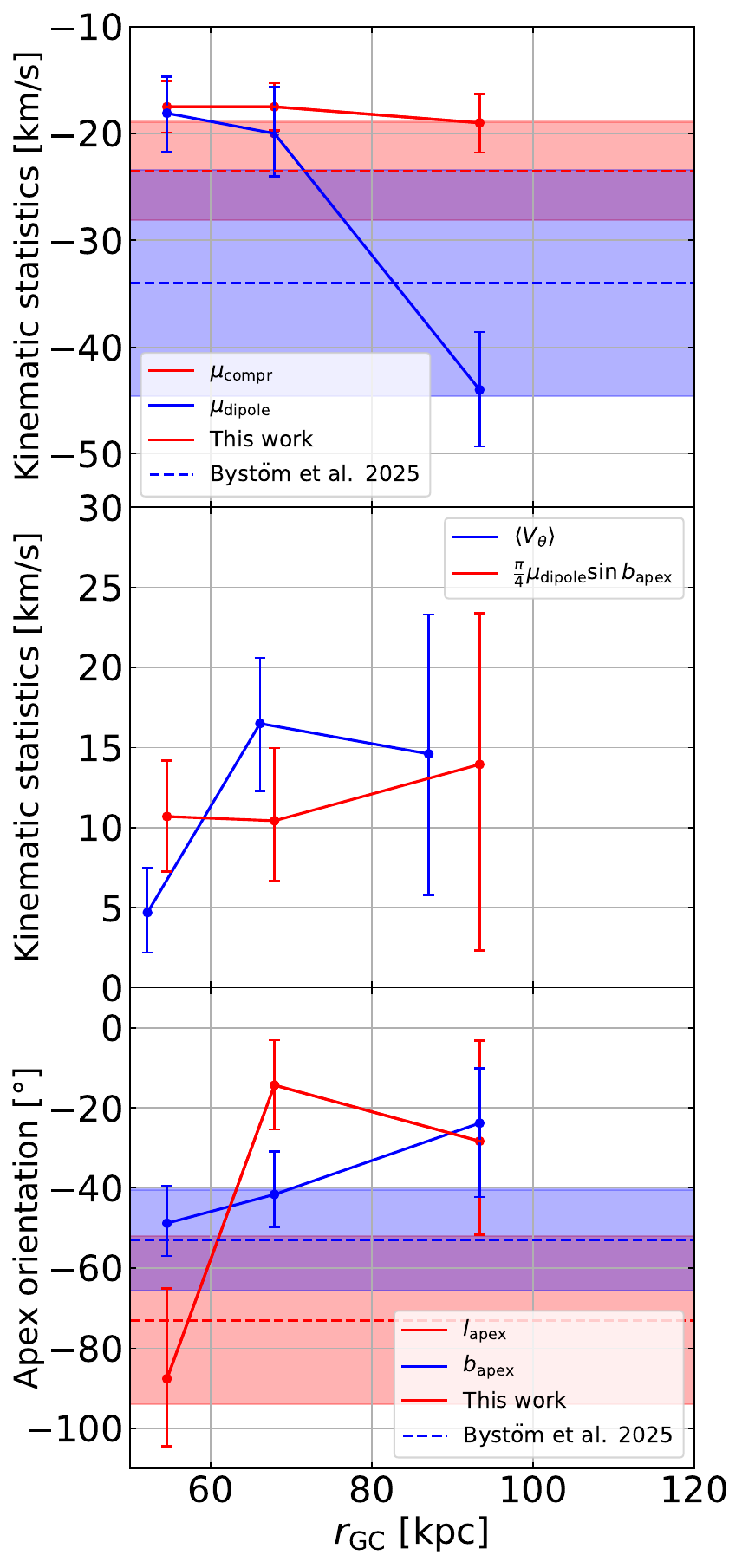}%
\end{center}
\caption{{\bf Top:} The contracting motion ($\mu_\mathrm{compr}$) and reflex motion ($\mu_\mathrm{dipole}$) of outer outer stelalr halo. 
{\bf Middle:} The mean polar velocity profile measured by the single-component model (blue line, see Section~\ref{sec:single_component_results}) and predicted by Equation~\ref{equ:predicted_v_theta} (red line).
{\bf Bottom:} The apex direction of the dipole velocity field at different radii. The red and blue lines represent the $l_\mathrm{apex}$ and $b_\mathrm{apex}$, respectively. 
The solid lines show the best-model in this work, with the errorbars present the 1$\sigma$ uncertainties. The dashed lines represent the results in \cite{2025MNRAS.542..560B}, with the shaded regions showing the 1$\sigma$ uncertainties.}
\label{fig:miu_profile}
\end{figure}

Figure~\ref{fig:LMC_overall_perturbation} illustrates the mean radial velocity $\langle V_r\rangle$ for the observation (left panel) and best-fit model (right panel) in Mollweide projection. The observation covers a distance range of $50~\mathrm{kpc}<r_\mathrm{GC}<120~\mathrm{kpc}$. The white dots in the two panels denote the position of the apex of the dipole velocity field, and red pentagrams are in proximity to the current location of the LMC. 

In the left panel of Figure~\ref{fig:LMC_overall_perturbation}, the observation is pixelized by the Python package \texttt{healpy} \citep{healpy}. We only retain the pixels that contain at least 15 stars, and the mean radial velocity assigned to each pixel is the inverse variance-weighted mean of all measurements falling within that pixel.\footnote{Here, we do not fit the mean velocity within a single pixel using a single Gaussian, as the small number of stars in many pixels would yield poorly constrained parameters.}. The left panel of Figure~\ref{fig:LMC_overall_perturbation} reveals a striking hemispheric asymmetry: pixels in the northern hemisphere exhibit both positive and negative $\langle V_r\rangle$ values that largely cancel, yielding no significant mean signal, whereas almost every southern pixel is negative, producing a coherent infalling signature.  This behaviour is fully consistent with the dipole pattern shown in Figure~\ref{fig:mean_vr_profile}. The right panel of Figure~\ref{fig:LMC_overall_perturbation} confirms this agreement: the southern sky exhibits strongly negative $\langle V_r\rangle$ values, with the strongest signal close to the current position of the LMC, while the northern sky shows no discernible mean radial velocity signature.

We further subdivide the K giant sample into three distinct radial bins:  $50~\mathrm{kpc}<r_\mathrm{GC}<60~\mathrm{kpc}$, $60~\mathrm{kpc}<r_\mathrm{GC}<80~\mathrm{kpc}$, and $80~\mathrm{kpc}<r_\mathrm{GC}<120~\mathrm{kpc}$ to investigate the dependence of $\mu_\mathrm{compr}$, $\mu_\mathrm{dipole}$ and the position of apex on Galactocentric distance.

The top panel of Figure~\ref{fig:miu_profile} presents the $\mu_\mathrm{compr}$ and $\mu_\mathrm{dipole}$ profiles as a function of distance. And Table~\ref{table:LMC_results} lists the best-fit parameters and $1\sigma$ uncertainties across various radial bins. In comparison, we also plot the best-fit model in \cite{2025MNRAS.542..560B}.
We find that: (1) $\mu_\mathrm{compr}$ exhibits no significant dependence on Galactocentric distance, indicating that the stellar halo in the range $50~\mathrm{kpc}<r_\mathrm{GC}<120~\mathrm{kpc}$ is undergoing a stable and sustained contraction. (2) the absolute value of $\mu_\mathrm{dipole}$ increases with distance, rising from $\sim-19~\mathrm{km~s^{-1}}$ in the $50~\mathrm{kpc}<r_\mathrm{GC}<60~\mathrm{kpc}$ bin to $\sim-44~\mathrm{km~s^{-1}}$ in the $80~\mathrm{kpc}<r_\mathrm{GC}<120~\mathrm{kpc}$ bin. The increasing $\mu_\mathrm{dipole}$ is consistent with \cite{2024MNRAS.527..437V}, as the predicted amplitude of the reflex motion is increasing with distance (see orange dash-dotted line in Figure~\ref{fig:mean_vr_profile}). Similarly, \cite{2025ApJ...988..156C} also states that the amplitude of the LMC-induced reflex motion increases monotonically with Galactocentric radius, (see Table~1 and Table~2 in this paper), rising from $\sim6~\mathrm{km~s^{-1}}$ (at $\sim$~40~kpc) to $\sim47~\mathrm{km~s^{-1}}$ (at $\sim$~120~kpc), in agreement with our results.

The middle panel of Figure~\ref{fig:miu_profile} then presents the comparison between the observed mean polar velocity fit by the single-component model (blue line, see Section~\ref{sec:single_component_results}) and the values directly deduced from $\mu_\mathrm{dipole}$ (red line). Based on \cite{2025MNRAS.544.2434S} (see their Equation~17): the mean polar velocity of the outer stellar halo stars scales linearly with the amplitude of the LMC-induced dipole velocity field ($\mu_{\text{dipole}}$), following:
\begin{align}
  \langle V_\theta \rangle = \frac{\pi}{4} \mu_{\text{dipole}} \sin b_\mathrm{apex}.
\label{equ:predicted_v_theta}
\end{align}
This link arises from the reflex motion of the MW in response to the LMC's gravitational pull, making polar velocity a direct tracer of the LMC-driven kinematic dipole. Figure~\ref{fig:miu_profile} shows that the observed mean polar velocity is consistent with the values deduced from our measured $\mu_\mathrm{dipole}$ and Equation~\ref{equ:predicted_v_theta}, confirming the linear link between polar velocity and the LMC-induced dipole velocity field in the outer stellar halo.

The bottom panel of Figure~\ref{fig:miu_profile} shows that the position of the apex of the dipole velocity field also evolves with Galactocentric radius, which is also reported in \cite{2025ApJ...988..156C} (see Figure~7 in this paper). Specifically, the Galactic latitude of the apex ($b_\mathrm{apex}$) approaches the midplane (i.e., becomes closer to $b_\mathrm{apex}=0$) with increasing distance, while the Galactic longitude $l_\mathrm{apex}$ also exhibits a general trend of increasing with distance (i.e., becomes farther away than the current location of the LMC). The change in the apex position with distance reflects the time-evolving impact of LMC on the MW outer stellar halo upon its infall. The more distant stars give an apex position futher away from the current position of the LMC, but is more consistant with its past orbit. The innermost distance bin shows a better consistency with \cite{2025MNRAS.542..560B}. This suggests that the difference between our measurements and \cite{2025MNRAS.542..560B} is due to the fact that our sample contains more distant stars. Note the median distance of the BHB subsample beyond 50~kpc in \cite{2025MNRAS.542..560B} is about 60~kpc. In comparison, our median distance for the subsample beyond 50~kpc is close to 70~kpc, with more stars extending to even larger distances.

According to Figure~\ref{fig:mean_vr_profile} and Figure~\ref{fig:miu_profile}, we conclude that the apparent net radial motions observed in the northern and southern outer stellar haloes represent the superposition of a global inward contraction and the reflex motion induced by the LMC’s infall. The overall contracting motion shares a similar amplitude to the reflex motion, leading to a nearly-zero net radial motion in the northern stellar halo. 
Furthermore, the observed mean polar velocity in the outer stellar halo can be well described by a linear relationship with the amplitude of the dipole velocity field.
However, the physical driver of this contracting motion remains elusive. Although the N-body simulations of \citet{2024MNRAS.527..437V} likewise imply an overall outer-halo contraction, as they find a smaller net radial signal in the north (see Figure~\ref{fig:mean_vr_profile}, but still greater than the observation) than in the south.

\begin{table}[ht]
\centering
\caption{The best-fit parameters of Equation~\ref{equ:LMC_mean_velocity_field} in several distance bins.}
\renewcommand{\arraystretch}{1.6}
\scriptsize      
\setlength{\tabcolsep}{3.5pt}  
\begin{tabular}{@{}lccccc@{}}
\toprule
$r_\mathrm{GC}$ & $\mu_\mathrm{compr}$ & $\mu_\mathrm{dipole}$ & $l_\mathrm{apex}$ & $b_\mathrm{apex}$ & $\sigma_r$\\
kpc & $\mathrm{km~s^{-1}}$ & $\mathrm{km~s^{-1}}$ & & & $\mathrm{km~s^{-1}}$ \\
\midrule
$[50,120]$ & -$17.9_{-1.9}^{+1.9}$ & -$24.0_{-3.8}^{+3.7}$ & -$34.8_{-7.8}^{+7.9}$ & -$3787_{-7.0}^{+5.9}$ & $149.1_{-0.9}^{+1.0}$\\
$[50,60)$ & -$17.5_{-2.4}^{+2.4}$ & -$18.1_{-3.4}^{+3.6}$ & -$87.6_{-22.5}^{+16.9}$ & -$48.8_{-9.3}^{+8.2}$ & $144.3_{-1.2}^{+1.1}$\\
$[60,80)$ & -$17.5_{-2.2}^{+2.2}$ & -$20.0_{-4.4}^{+4.0}$ & -$14.3_{-11.2}^{+11.1}$ & -$41.6_{-10.7}^{+8.2}$ & $148.8_{-1.0}^{+1.0}$\\
$[80,120]$ & -$19.0_{-2.7}^{+2.8}$ & -$44.0_{-5.4}^{+5.3}$ & -$28.3_{-25.1}^{+23.4}$ & -$23.8_{-13.7}^{+18.4}$ & $148.4_{-1.2}^{+1.3}$\\
\midrule
\bottomrule
\end{tabular}
\label{table:LMC_results}
\end{table}

\section{Discussions: Possible Mechanism of the Net Motion of the Stellar Halo}
\label{sec:discussion}

In this Section, we summarize a few possible mechanisms that may induce the observed net motion (i.e., net azimuthal rotation in the inner region and net contracting motion in the outer part) of the stellar halo.

Assuming the satellites fell in the MW with random orbits, the overall contribution of the angular momentum should be around zero \citep{2015ApJ...801...98S} from minor mergers, leading to no net motion in radial, azimuthal, and polar velocities. 
However, in reality, the stellar halo has net rotation and even contracting motion. Section~\ref{sec:single_component_results} and Section~\ref{sec:doubel_components_results2} have shown that the inner stellar halo exhibits net azimuthal rotation, and the outer stellar halo presents net motion in both polar and radial velocities (see Figure~\ref{fig:single_component_result} and Figure~\ref{fig:double_component_results}).

The net azimuthal rotation for the stellar halo might reflect the co-formation of the Galactic disk and the stellar halo \citep{2018MNRAS.478..611B}.
According to the N-body simulations for those barred galaxies \citep{2007MNRAS.377.1569A,2010MNRAS.406.2386M,2013MNRAS.429.1949A,2018MNRAS.476.1331C,2022ApJ...940..175K}, a rotating bar in the Galactic center can transfer angular momentum to the dark matter and stellar halo, leading to an inner stellar halo prograde with the disk. In our analysis, we find stronger rotation in the MP population. The older MP population may have had more time to interact with the MW bar and therefore has a higher rotation amplitude. However, this cannot explain the spin of the MW stellar halo at greater distances.

The net motion of the stellar halo could also be directly related to the anisotropic accretion of satellite galaxies, not only including the minor mergers, but also including the major merger events, such as the GSE. As the GSE remnant contributes a larger fraction of stars in the stellar halo, the observed net motion may reflect the specific kinematics of the progenitor itself. However, the underlying reason for the stronger net azimuthal rotation of the MP population compared to the MR population remains unclear. The difference might be due to the different merger history of the two populations.

Interestingly, \cite{2017MNRAS.470.1259D} investigated the spin of 30 Auriga stellar halos and reported that the Auriga stellar halo spins are significantly larger than the spin of our MW stellar halo. Moreover, old halo stars show weaker rotations in Auriga. These likely indicate that our MW could have experienced a quiescent merger history in the past, consistent with quite a few previous studies based on other observational evidence \citep[e.g. Tao et al., in preparation][]{2015MNRAS.450.2874R,2020MNRAS.494.5936F,2023MNRAS.520.6091D}, before LMC falling in. More detailed investigations based on numerical simulations will be presented in Batrakov et al., in preparation.

The observed inward motion of the outer stellar halo might arise because stripped LMC material increases the gravitational potential at smaller radii, effectively pulling halo stars inward \citep{2021NatAs...5..251P,2024MNRAS.531.3524Y,2025MNRAS.542..560B}. However, if the LMC is currently on its first orbit, it might be hard to imagine it to significantly deepen the potential in inner halos given its current distance of 50~kpc. However, \cite{2023MNRAS.521.4936K} reported that the radius of LMC subhalo is about 50~kpc, and thus its extended subhalo boundary may have reached smaller radii than the LMC itself. 

Moreover, \cite{2025MNRAS.542..560B} also suggested that the inward motion could be explained by some unrelaxed substructures and the local wake of LMC, as the residual accretion debris could still carry net infalling motion, and the dynamical friction wake behind the LMC adds extra negative radial velocity along its orbit (i.e., orbital decay, see Section~5.2 in \cite{2025MNRAS.542..560B} for details). Inward motion due to unrelaxed substructures is not surprising and could be universally existent, because the host halo keeps accreting material and growing. Such inward motion, however, could fail to be identified in isolated MW-LMC system simulations.

\section{Conclusions}
\label{sec:concl}

From the DESI stellar surveys DR2, we select a halo K giant sample within $3~\mathrm{kpc}<r_\mathrm{GC}<160~\mathrm{kpc}$ and study the mean metallicity, mean velocity, velocity dispersion, and velocity anisotropy profiles as a function of Galactocentric distance using GMM. We also measure the reflex motion and contracting motion of the outer halo with a dipole velocity field model. 

By modelling the velocity and metallicity distributions using all halo K giants ($3~\mathrm{kpc}<r_\mathrm{GC}<160~\mathrm{kpc}$), we identify that the stellar halo is composed of two distinct and nearly equally contributed components: the metal-rich (MR) and the metal-poor (MP) populations. These two components are distinct in both velocity distribution and metallicity:\\
(1) The median metallicity ($\mu_\mathrm{[Fe/H]}$) of the MR component is about $\sim-1.15~\mathrm{dex}$ and the MP component has $\mu_\mathrm{[Fe/H]}\sim-1.58~\mathrm{dex}$. We also confirm that the MR population displays a markedly narrower metallicity dispersion ($\sigma_\mathrm{[Fe/H]}^\mathrm{MR}\sim0.17~\mathrm{dex}$) than its MP counterpart ($\sigma_\mathrm{[Fe/H]}^\mathrm{MP}\sim0.43~\mathrm{dex}$), consistent with \cite{2024ApJ...974..167Z} and \cite{yr1rrlyrae}. \\
(2) The MR population exhibits noticeably smaller azimuthal and polar velocity dispersion than the MP population, indicating a more radial velocity anisotropy parameter than the MP population ($\beta\sim0.94$ for the MR population and $\beta\sim0.46$ for the MP population). Following \cite{yr1rrlyrae}, we suggest the MR population to be highly correlated with the GSE merger events, due to its high velocity anisotropy.\\
(3) Both populations exhibit a net azimuthal rotation. However, the MP component displays a higher rotation amplitude than that of the MR component. The discrepancy might be due to the different merger history of the two populations.

We further model the velocity and metallicity distributions within discrete Galactocentric annuli and explore the mean velocity, velocity dispersion, and velocity anisotropy profiles as a function of distance. We find that:\\
(1) The fractional contribution of the MR component exhibits a non-trivial radial dependence similar to \cite{yr1rrlyrae}, contributing more than 40\% of the total stellar halo even beyond $r_\mathrm{GC}>50~\mathrm{kpc}$. This result is consistent with the findings of \citet{2023ApJ...951...26C}, who argued that debris from the GSE merger event remains detectable at similarly large Galactocentric distances.\\
(2) The MR population exhibits radial-dominated orbits with $\beta_\mathrm{MR}>0.8$ beyond 10~kpc, indicating that the MR population is mainly contributed by the GSE merger event, whereas the MP population presents a velocity profile decreasing from $\beta_\mathrm{MP}\sim$0.5 ($\sim$~10~kpc) to -0.5 ($\sim$~85~kpc), reflecting that the MP population could be shaped by multiple minor merger events. \\
(3) Both populations exhibit a net azimuthal rotation over the distance range of 10~kpc to 30~kpc, and we suggest that the net azimuthal rotation is intrinsic to both populations.

With the dipole velocity model, we re-identify the reflex motion of the MW due to the infall of the LMC, and the inward motion (contracting motion) of the outer stellar halo. 
We find that:\\
(1) Similar to \cite{2025MNRAS.542..560B}, the southern stellar halo exhibits a stronger net radial motion ($\langle V_r\rangle$) than the northern stellar halo, in contrast to the prediction from N-body simulation \citep{2024MNRAS.527..437V}. The discrepancy could be explained by a contracting motion of the outer stellar halo, i.e, the outer stellar halo is undergoing a contraction, leading to a weaker $\langle V_r\rangle$ in the northern sky.\\
(2) We find that the amplitude of the reflex motion shares a similar amplitude with the contracting motion at $50~\mathrm{kpc}<r_\mathrm{GC}<60~\mathrm{kpc}$, and is increasing with distance, consistent with the prediction from \cite{2024MNRAS.527..437V}. The amplitude of the contracting motion does not show dependence on distance.\\
(3) We confirm the linear relationship between the mean polar velocity and the LMC-induced dipole velocity field in the outer stellar halo predicted by simulations \citep{2025MNRAS.544.2434S}.

We conclude that our results provide a comprehensive characterization of the chemo-dynamical signatures of the stellar halo, revealing distinct signatures of multiple merger events and the ongoing influence of the LMC. The identified MR and MP components, with their unique kinematic and chemical properties, reflect the complex assembly history of the Milky Way. The observed azimuthal rotation and radial contraction highlight the dynamic interplay between internal halo processes and external perturbations. Future work will focus on refining these models with additional data and exploring the implications for the Milky Way’s formation and evolution.

The measurements presented in this paper can be accessed at \url{https://doi.org/10.5281/zenodo.19176800}, which contains the data points for the figures presented in this work.

\acknowledgments
This work is supported by NSFC (12573022, 12595312, 12273021), the National Key R\&D Program of China (2023YFA1605600, 2023YFA1605601), 111 project (No.\ B20019), and the Office of Science and Technology, Shanghai Municipal Government (grant Nos. 24DX1400100, ZJ2023-ZD-001). We thank the sponsorship from Yangyang Development Fund. The computations of this work are carried on the National Energy Research Scientific Computing Center (NERSC) and the Gravity supercomputer at the Department of Astronomy, Shanghai Jiao Tong University. SK acknowledges support from the Science \& Technology Facilities Council (STFC) grant ST/Y001001/1. 

WW is grateful for useful discussions with Xiaoting Fu, Maosheng Xiang, 
Xiangxiang Xue, Hao Tian, Jundan Nie, Hongliang Yan, Jie Wang and Wenda Li. 

This material is based upon work supported by the U.S. Department of Energy (DOE), Office of Science, Office of High-Energy Physics, under Contract No. DE–AC02–05CH11231, and by the National Energy Research Scientific Computing Center, a DOE Office of Science User Facility under the same contract. Additional support for DESI was provided by the U.S. National Science Foundation (NSF), Division of Astronomical Sciences under Contract No. AST-0950945 to the NSF’s National Optical-Infrared Astronomy Research Laboratory; the Science and Technology Facilities Council of the United Kingdom; the Gordon and Betty Moore Foundation; the Heising-Simons Foundation; the French Alternative Energies and Atomic Energy Commission (CEA); the National Council of Humanities, Science and Technology of Mexico (CONAHCYT); the Ministry of Science, Innovation and Universities of Spain (MICIU/AEI/10.13039/501100011033), and by the DESI Member Institutions: \url{https://www.desi.lbl.gov/collaborating-institutions}. Any opinions, findings, and conclusions or recommendations expressed in this material are those of the author(s) and do not necessarily reflect the views of the U. S. National Science Foundation, the U. S. Department of Energy, or any of the listed funding agencies.

The authors are honored to be permitted to conduct scientific research on I'oligam Du'ag (Kitt Peak), a mountain with particular significance to the Tohono O’odham Nation.
DESI Legacy Survey with DESI Standard Acknowledgements

For the purpose of open access, the author has applied a Creative
Commons Attribution (CC BY) licence to any Author Accepted
Manuscript version arising from this submission. 

\clearpage

\appendix
\section{Effect of distance uncertainties}
\label{app:error_on_model}

Owing to the negative radial density profile gradient in the MW stellar halo, distance uncertainties may scatter more inner halo stars outwards, contaminating the outer halo sample and biasing the velocity anisotropy and MR fraction profiles at larger distances. This appendix briefly discusses the impact of K giant distance uncertainties on our results.

To quantify the impact of distance uncertainties on our results, we first generate ideal mock data and compare the change after including distance uncertainties. We adopt a simplified setup where the stellar halo is assumed to be spherical, and its radial density profile follows a triple power law with parameters taken from \cite{li2025milkywaystellarhalo}.
We further assume the velocity anisotropy profile of the MR population is constant ($\beta_\mathrm{MR}=0.9$) based on our measurements and other previous measurements presented in the main text\footnote{The assumption may not be realistic, as the nearly constant velocity anisotropy for the MR stellar halo might have already been affected by distance errors, but here we mainly focus on investigating the fractional changes in our measurements due to distance uncertainties, and thus the assumed form does not matter much.}. In contrast, that of the MP population is taken from our measurements presented in Table~\ref{table:double_component_results}. We additionally adopt the fraction of the MR population ($P_\mathrm{MR}$) measured in \cite{yr1rrlyrae} (see Figure 16 in this paper, adopting $p_\mathrm{GSE}>0.5$ cut). For each mock star at a given Galactocentric radius, we perform Monte Carlo sampling to assign membership to either the MR or MP population: stars assigned to the MR population are given a constant velocity anisotropy, whereas the velocity anisotropy of MP population stars is determined via linear interpolating our measured MP data points. In the end, we perturb the distances of the mock stars according to 18\% and 30\% relative distance uncertainties. 

\begin{figure}
\begin{center}
\includegraphics[width=0.9\textwidth]{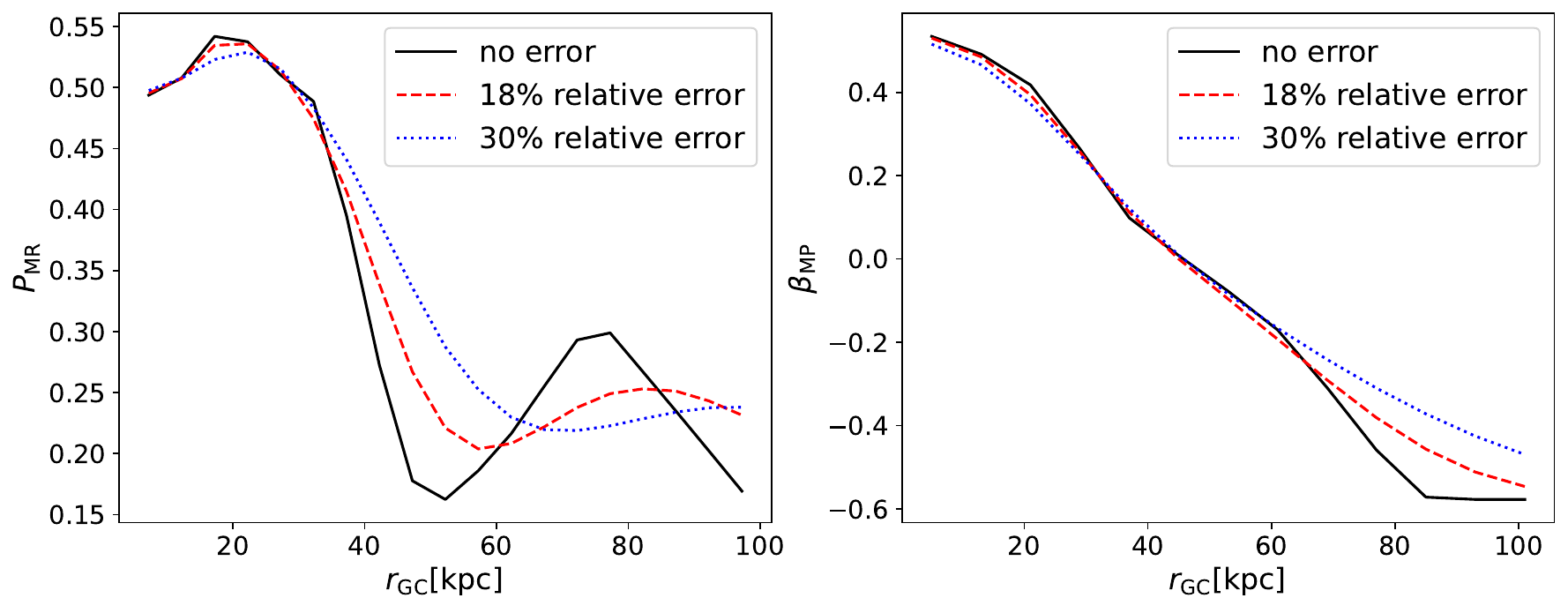}%
\end{center}
\caption{Impact of distance uncertainties on the MR population fraction (left panel) and MP population velocity anisotropy (right panel) profiles. Black solid lines are reference measurements, red dashed/blue dotted lines correspond to 18\%/30\% relative distance uncertainties, respectively.}
\label{fig:distance_uncertainty}
\end{figure}

Figure~\ref{fig:distance_uncertainty} illustrates how distance uncertainties introduce biases to our results. The red dashed line corresponds to an 18\% relative distance uncertainty, the typical uncertainty for \texttt{SpecDis2} derived from external validations using member stars in GCs \citep{2021MNRAS.505.5957B} and dwarf galaxies \citep{2022ApJ...940..136P}. For comparison, the blue dotted line shows results with a 30\% relative distance uncertainty. 

The left panel of Figure~\ref{fig:distance_uncertainty} shows the impact of distance uncertainties on the MR fraction profile: the black solid line denotes the reference measurement from \cite{yr1rrlyrae}. The red dashed and blue dotted lines present the perturbed profiles, which overestimate the MR population fraction between 40~kpc and 60~kpc, owing to the outward scattering of inner MR stars. The blue line exhibits a more significant overestimation than the red solid line, consistent with its larger relative distance error.

The right panel of Figure~\ref{fig:distance_uncertainty} illustrates the effect of distance uncertainties on the MP velocity anisotropy profile. Given our assumption of a constant velocity anisotropy for the MR population, its profile is immune to distance uncertainties, and we therefore omit the MR population profile from the plot. 
The black solid line denotes the reference MP velocity anisotropy profile. The red dashed and blue dotted lines represent the perturbed profile due to 18\% and 30\% distance uncertainties. Both the red dashed and blue dotted lines are consistent with the distance error-free black line between 30 and 60~kpc. There are some slight underestimates within 30~kpc, and more prominent overestimates beyond 60~kpc after including distance uncertainties. 

We conclude that distance uncertainties bias the observed MR fraction profile and MP population velocity anisotropy profile. Specifically, such uncertainties lead to overestimates in both the outer stellar halo MR fraction and the distance where $P_\mathrm{MR}$ ceases decreasing and starts increasing. Moreover, distance uncertainties flatten the velocity anisotropic profiles at larger radii. However, the fractional change is not large enough to violate our conclusions in the main text, even with the 30\% of distance error.

\bibliography{master}{}
\bibliographystyle{aasjournal}

\end{document}